\documentclass[conference]{IEEEtran}
\usepackage{amsmath,amssymb,amsfonts}
\usepackage{amsthm}
\usepackage{bm}
\let\labelindent\relax
\usepackage{enumitem}
\usepackage{graphicx}
\usepackage{balance}
\usepackage{booktabs}
\usepackage[framemethod=TikZ]{mdframed}
\usepackage{caption}
\usepackage{multirow}
\usepackage{subcaption}
\usepackage{xcolor}
\usepackage{xcolor,colortbl}
\usepackage{amssymb}
\usepackage{pifont}
\usepackage{algorithm}
\usepackage{algpseudocode}
\usepackage{siunitx}
 \usepackage{threeparttable}
\usepackage{tcolorbox}
\usepackage[hidelinks]{hyperref}
\newtheorem{definition}{Definition}

\usepackage{tikz}
\newcommand{\pie}[1]{%
\begin{tikzpicture}
 \draw (0ex,0ex) circle (1ex);
 \fill (0ex,-1ex) arc (-90:(#1-90):1ex) -- (0ex,-1ex) -- cycle;
\end{tikzpicture}%
}

\newcommand{\rect}[1]{%
\begin{tikzpicture}
 \draw (-1ex,-1ex) rectangle (1ex,1ex);
 \fill (#1ex,-1ex) rectangle (1ex,1ex);
\end{tikzpicture}%
}

\usepackage{xspace}
\def\eg{\emph{e.g.,}\xspace}

\def\ie{\emph{i.e.,}\xspace}
\def\etal{\emph{et al.}\xspace}

\ifCLASSINFOpdf
\else
\fi
\begin{document}
\title{A Duty to Forget, a Right to be Assured? Exposing Vulnerabilities in Machine Unlearning Services}
\author{\IEEEauthorblockN{Hongsheng Hu\IEEEauthorrefmark{1},
Shuo Wang\IEEEauthorrefmark{4}\IEEEauthorrefmark{1}, Jiamin Chang\IEEEauthorrefmark{2}\IEEEauthorrefmark{1}, Haonan Zhong\IEEEauthorrefmark{2}\IEEEauthorrefmark{1}, Ruoxi Sun\IEEEauthorrefmark{1}, \\Shuang Hao\IEEEauthorrefmark{3}, Haojin Zhu\IEEEauthorrefmark{4}, and Minhui Xue\IEEEauthorrefmark{1} 
}
\IEEEauthorblockA{\IEEEauthorrefmark{1}CSIRO's Data61, Australia}
\IEEEauthorblockA{\IEEEauthorrefmark{2}University of New South Wales, Australia}
\IEEEauthorblockA{\IEEEauthorrefmark{3}University of Texas at Dallas, USA} \IEEEauthorblockA{\IEEEauthorrefmark{4}Shanghai Jiao Tong University, China}}

\IEEEoverridecommandlockouts
\makeatletter\def\@IEEEpubidpullup{6.5\baselineskip}\makeatother
\IEEEpubid{\parbox{\columnwidth}{
    Network and Distributed System Security (NDSS) Symposium 2024\\
    26 February - 1 March 2024, San Diego, CA, USA\\
    ISBN 1-891562-93-2\\
    https://dx.doi.org/10.14722/ndss.2024.24252\\
    www.ndss-symposium.org
}
\hspace{\columnsep}\makebox[\columnwidth]{}}

\maketitle

\begin{abstract}
The right to be forgotten requires the removal or ``unlearning'' of a user's data from machine learning models. However, in the context of Machine Learning as a Service (MLaaS), retraining a model from scratch to fulfill the unlearning request is impractical due to the lack of training data on the service provider's side (the server). 
Furthermore, approximate unlearning further embraces a complex trade-off between utility (model performance) and privacy (unlearning performance). In this paper, we try to explore the potential threats posed by unlearning services in MLaaS, specifically over-unlearning, where more information is unlearned than expected. We propose two strategies that leverage over-unlearning to measure the impact on the trade-off balancing, under black-box access settings, in which the existing machine unlearning attacks are not applicable. The effectiveness of these strategies is evaluated through extensive experiments on benchmark datasets, across various model architectures and representative unlearning approaches. Results indicate significant potential for both strategies to undermine model efficacy in unlearning scenarios. This study uncovers an underexplored gap between unlearning and contemporary MLaaS, highlighting the need for careful considerations in balancing data unlearning, model utility, and security.
\end{abstract}

\section{Introduction}
Deep Neural Networks (DNN) models are often trained on large amounts of data, including personal data~\cite{de2021critical}. 
Nevertheless, the General Data Protection Regulation (GDPR)~\cite{rosen2011right} and the California Consumer Privacy Act (CCPA)~\cite{pardau2018california}, \ie \textit{Right to be Forgotten}, enforce the service providers to necessitate removal of users' training data when requested by individuals due to privacy regulation compliance. 
To practically grant the right to be forgotten, machine unlearning techniques~\cite{bourtoule2021machine} are developed, aiming to protect users' privacy by removing the contribution of a data sample or several data samples from a trained ML model on request or after a particular timescale.
However, machine unlearning on deep models is still in its infancy. Deleting data from a database~\cite{cao2015towards} can be relatively straightforward and intuitive, while it is much more complicated in the case of deep models because of the model complexity and the randomness in the training algorithms~\cite{thudi2022unrolling}. 
A commonly adopted unlearning approach is retraining, \ie retraining the model from scratch where the data to be unlearned are removed. However, retraining is almost always associated with heavy computational cost, especially when models are deep models with millions or even billions of parameters.

\begin{figure}[t]
\centering
\includegraphics[width=1.0\linewidth]{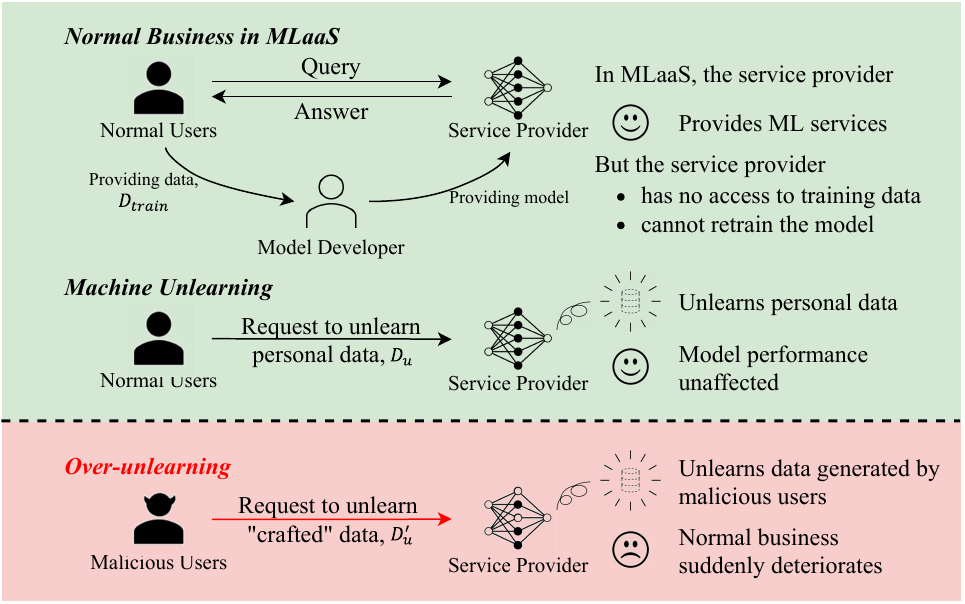}
\caption{An overview of the over-unlearning threat in machine unlearning as a service.}
\label{fig:threats}
\end{figure}

The trend of Machine Learning as a Service (MLaaS) has gained significant momentum in recent years~\cite{imarc2023machine}. 
Implementing deep models on the cloud platform as API services enjoys several advantages, such as privacy enhancement (splitting the storage of training data from service provider), accessibility, cost-effectiveness, and scalability~\cite{shmueli2023machine}. 
One successful example is the case of Lufthansa Technik using the MLaaS of Google Cloud AutoML~\cite{mlaas}. 
Neglecting to eliminate data from the deployed models timely may lead to punitive measures, reinforcing the importance of effective machine unlearning strategies in cloud-based model management. For example, UK’s data regulator, the Information Commissioner’s Office (ICO), has issued guidance on the data protection implications of using AI and generative models~\cite{uk}. The U.S. Federal Trade Commission (FTC) made the cloud storage application Ever delete both user data and any deployed models trained on users' data in 2021~\cite{federal2021ftc}.
However, to achieve data removal on deep models in MLaaS, the feasibility of prevailing efficient retraining methods is often compromised due to the absence of the server's direct access to the original training dataset. Conducting retraining procedures on a local scale and subsequent redeployment also incurs supplementary costs and induces delays. Ideally, the unlearning procedure could be hosted by the cloud as well.
A promising industrial paradigm of data removal from the cloud database is exemplified by Google Analytics 4, involving the utilization of the User Deletion API~\cite{googledele}.
Fortunately, some approximate unlearning approaches directly modify the model parameters and do not require the original training dataset to be integrated as an Unlearning API, which can greatly facilitate the server to provide machine unlearning services. 
Thus, similarly, with the Unlearning API, a user who once contributed data to train the model can freely ask the service provider for the deletion of her or his data on the model if she or he wishes to opt out, as shown in \autoref{fig:threats}. 

\noindent \textbf{Research Gap.} 
Current machine unlearning methodologies have been developed and assessed within a local development context, wherein the developer has comprehensive access to both the model and its training data. However, the advent of Machine Learning as a Service (MLaaS) imposes limitations on the availability of training data and resources, rendering most existing unlearning approaches~\cite{bourtoule2021machine,golatkar2020eternal,kim2022efficient,tarun2023fast,baumhauer2022machine} to be futile. 
Furthermore, MLaaS transitions the model from a local environment to a public deployment setting, introducing the potential for untrustworthy or even malicious users. These users may seek to carry out compromising activities through unlearning requests, reducing the practicability to implement the machine unlearning process in the MLaaS context.
\textit{Given some approximate unlearning methods can be applied in the MLaaS, however, none of the existing works investigate what types of risks malicious users can bring to the server.}

\noindent \textbf{Research Question.} 
From the standpoint of the MLaaS server, a balance must be struck between sustaining normal business services (primary tasks) and accommodating machine unlearning services (secondary tasks). Catering to unlearning requests from users inevitably diminishes the model's utility, as the contributions made by the user's data are excised.
As such, the deployment of the unlearning process within the MLaaS platform hinges on the assurance of equilibrium between model fidelity and unlearning effectiveness. 
In this context, the research question is: \textit{Is it feasible for a user to compromise the normal business of the server by requesting machine unlearning services in MLaaS? And how easily can the user achieve the compromise?}
{Unfortunately, existing research either tends to concentrate on fooling the model, or neglects to evaluate the trade-off entirely, hence falling short in addressing the research question due to their inability to satisfy the constraints of MLaaS settings (detailed in Section~\ref{sec:pre}).}

\noindent \textbf{Motivation.} To answer this question, we conduct the first investigation to identify the potential threats that can be associated with machine unlearning services under MLaaS scenarios. We identify the major threat that a user can pose to the server: \textit{over-unlearning}. 
Specifically, as depicted in \autoref{fig:threats} and \autoref{fig:unlearning}, over-unlearning represents that the information the server's model unlearned exceeds what it ought to unlearn by manipulating the unlearned samples to contain more knowledge than expected. 

To mimic the malicious users' behaviors and measure the feasibility of potential risks, we present two strategies that can be leveraged for the manipulation to achieve over-unlearning, while only granting the user black-box access to the model of the server. Specifically, over-unlearning is instigated by deliberately \textit{blending} additional samples from a disparate task into the original unlearned samples through meticulous sample-level manipulation.  
When the server unlearns the blended data, the model will remove the additional information about that particular task.
Furthermore, we propose an advanced over-unlearning approach by \textit{pushing} the samples close to the decision boundary via pixel-wise manipulation. We consider that the user can intentionally augment the unlearning effect of the unlearned data on the server's model by moving the unlearned data to the decision boundary of the model. Compared to the original unlearned data, the modified unlearned data become more informative to the model. Thus, by uploading the modified data instead of the original data for the server to unlearn, the model unlearns more information than the case of unlearning the original data, which achieves the goal of over-unlearning. 
By answering this question, we try to provide an evaluation pipeline toward the deployability of machine unlearning in real-world MLaaS implementing scenarios. 

\noindent \textbf{Contributions.} Our contributions are summarized as follows:
\begin{itemize}[leftmargin=*]
\item This paper is the \textit{first} to investigate the real-world machine unlearning service pipeline and to identify potential threats related to machine unlearning services in the MLaaS environment. It illuminates the risk of over-unlearning, which could significantly compromise the model utility of the server when the malicious unlearning request is submitted.

\item We identify two novel strategies that allow a user to achieve over-unlearning, with only black-box access to the server's model, making them easily applicable in real-world MLaaS environments.

\item Extensive experiments are conducted using benchmark datasets, across various model architectures and representative unlearning approaches. The findings validate the effectiveness of the proposed strategies in inducing over-unlearning. 

\item An extensive ablation study is performed to evaluate the effectiveness of the proposed strategies in different settings. The results demonstrate the effectiveness of the strategies across various settings, emphasizing their potential impact and the importance of mitigating such risks in MLaaS environments.

\end{itemize}

\section{Related Work and Threat Model}\label{sec:pre}
In this section, we first introduce recent studies in machine unlearning domain and then describe the threat model of machine unlearning service particularly in MLaaS scenarios.

\subsection{Related Work}
Machine unlearning is motivated by a variety of reasons: \textit{i)} to comply with privacy regulations; \textit{ii)} the trained model itself needs to be updated or repaired. For example, a model trained on a poisoned dataset exhibits unexpected or undesirable behavior in response to benign inputs or deliberately crafted inputs. Thus, the model needs to be repaired to ensure its safety~\cite{cao2015towards}.
To achieve machine unlearning, a naive yet natural approach is retraining, \ie retraining the model on the training dataset with the unlearned data removed. However, retraining the model can result in expensive computation costs when the model or the training dataset is large-scale. For example, it shows that 34 days are needed to train OpenAI's {GPT-3}~\cite{brown2020language} even on 1,024 NVIDIA A100 GPUs~\cite{narayanan2021efficient}. To overcome the limitation of naively retraining while fulfilling the demand of machine unlearning, many unlearning approaches have been proposed recently, and they can be mainly divided into two categories: exact unlearning and approximate unlearning.

\noindent \textbf{Exact Unlearning.} Exact machine unlearning requires the model to be retrained from scratch, \ie retrain the model on the training dataset with the unlearned data removed with some skills to achieve efficiency. The advantage of exact unlearning is that it guarantees the impacts of the unlearned data are completely removed because the unlearned model is never trained on the unlearned data. To achieve efficiency, either the training dataset or the model architecture is carefully crafted before the training process. Specifically, the first line of work splits the training dataset into different non-overlap shards, where a constituent model is trained on each of the shards. Thus, when unlearning a data sample, only the constituent model trained on the shard containing the unlearned data is required to be retrained. Based on the intuition of dataset partition, Bourtoule~\etal~\cite{bourtoule2021machine} propose the unlearning approach of SISA that is generic to a variety of models with different architectures and complexities. Another line of work focuses on carefully designing the architecture of the model so that a data sample only contributes to part of the model. When unlearning a data sample, only the partial of the model that is influenced by the unlearned data is required to be retrained. Based on carefully designing the tree structure, Brophy and Lowd~\cite{brophy2021machine} propose data removal-enabled forests that support efficient unlearning of a data sample for random forests. Although exact unlearning has a perfect unlearning guarantee, the disadvantage is that it usually requires storing the training dataset, which may limit its applicability in many cases like MLaaS, where users' data is not allowed to be stored or the training dataset is deleted to comply with data regulations.

\noindent \textbf{Approximate Unlearning.} Approximate machine unlearning directly modifies the parameters of the trained model to obtain the unlearned model that approximates the model retrained from scratch. Approximate unlearning is usually achieved by updating the parameters of the trained model with a few numbers of iterations using the information calculated from the unlearned data. There are mainly two kinds of approximate unlearning methods, which are based on the influence function~\cite{cook1980characterizations} and the gradient of the loss function, respectively. Specifically, the first one~\cite{guo2020certified,izzo2021approximate} leverages the influence function to calculate the influence of the unlearned data on the model parameters so that the trained model can apply a Newton step to remove the influence for obtaining the unlearned model
The advantage of the first kind of approximate unlearning method is that one-step Newton update is enough for removing the contribution of the unlearned data. However, the drawbacks are also obvious: \textit{i)} it requires computing the inverse Hessian matrix of the loss function, which can be difficult for large-scale deep models; \textit{ii)} the training dataset might not be applicable in cases where it is not stored or deleted.

The second kind of unlearning methods~\cite{thudi2022unrolling,wu2020deltagrad,neel2021descent} calculates the gradients of the unlearned data contributed to the trained model during the training process. Then, to unlearn the data, the trained model is updated by adding back these gradients to approximate the model that is retrained from scratch. A state-of-the-art gradient-based unlearning method is proposed by Warnecke~\etal~\cite{warnecke2023machine}. The high-level intuition of this unlearning method is to overwrite the unlearned data from the trained model. An advantage of this method is that it only requires to access the unlearned data, which makes it practical in many cases especially in MLaaS where only the unlearned data is applicable. 

In addition to the two kinds of unlearning methods, fine-tuning has been widely used as an empirical unlearning baseline in existing works~\cite{golatkar2020eternal,golatkar2020forgetting,graves2021amnesiac,xu2023machine}. Fine-tuning first randomly selects incorrect labels and uses them to relabel the unlearned samples. Then, the trained model is fine-tuned on these relabeled unlearned data for several iterations for unlearning. The intuition is to confuse the model’s understanding of the unlearned sample so that it cannot output the correct prediction of the unlearned data. However, information of the unlearned data can still remain in the parameters of the unlearned model produced by fine-tuning.

In this study, we mainly focus on evaluating the risks
that may exist in the gradient-based approximate unlearning approaches as they are feasible in MLaaS but the retraining method is not.
Particularly, we involve fine-tuning as the empirical unlearning baseline and the approaches used in Warnecke~\etal~\cite{warnecke2023machine} as the state-to-the-art approximation-based unlearning baseline in our evaluation.

\subsection{Our Threat Model}
In this subsection, we describe the potential threat to machine unlearning that may exist particularly in MLaaS scenarios. To formalize, we let $\mathcal{D}_{\textrm{train}}=\mathcal{D}_{u} \cup \mathcal{D}_{r}$, where $\mathcal{D}_{u}$ is the dataset that contains the unlearned data and $\mathcal{D}_{r}$ is the dataset that contains the remaining data. Let $\bm{\theta}^*$ be the model trained on $\mathcal{D}_{\textrm{train}}$. Machine unlearning aims to obtain an unlearned model $\bm{\theta}_{u}^*$ by removing the contribution of $\mathcal{D}_{u}$ that has contributed during the training process to $\bm{\theta}^*$.

\noindent \textbf{MLaaS Scenario.} We assume a developer who, upon training a model with dataset $\mathcal{D}_{\textrm{train}}$, retains proprietary ownership of the resulting model. Subsequent to the training phase, the developer deploys the model on a server to offer MLaaS service for commercialization. The server has a test dataset $\mathcal{D}_{\textrm{test}}$ and works as an agent responsible for model deployment and maintenance, including monitoring the performance of the models and updating them. We call the user who has once contributed training data $\mathcal{D}_u \subset \mathcal{D}_{\textrm{train}}$ to train the model as an \textit{\textbf{authorized unlearning user:}} they can revoke the contribution of their data due to data protection regulations. Generally, different unlearning procedures may involve retraining and have different impacts on the quality of the model. Thus, to maintain the model with the compliance of regulations, the developer has agreed on a pre-selected unlearning method that can ensure the balance between model fidelity and unlearning efficacy with the server. When an authorized unlearning user decides to revoke, the server will fulfill the unlearning requests using the pre-selected unlearning method. 

Under the MLaaS scenarios, there are three entities involved in machine unlearning services: the model provider, the MLaaS server, and the users who utilize the model. The model provider transfers or delegates the high-utility model to the MLaaS server for professional business services as well as machine unlearning services, while the model users can utilize the model provided by the MLaaS server via APIs. For authorized unlearning users, they can also raise unlearning requests by uploading $\mathcal{D}_u$ to the server for deleting their data when they decide to opt-out. 
The knowledge of the developer, users, and server is summarized in  \autoref{tab:knowledge}.

\begin{table}[t]
\centering
\caption{Knowledge of dataset available for different entities in MLaaS.}
\resizebox{0.9\linewidth}{!}{%
\begin{tabular}{lccc}
\toprule
& $\mathcal{D}_{\textrm{train}}$ & $\mathcal{D}_{u}$ & $\mathcal{D}_{\textrm{test}}$\\
\midrule
Model Provider & \pie{360} & Unknown & \pie{360} \\
MLaaS Server & \pie{0} & \pie{0} & \pie{360} \\
Authorized Unlearning User & \pie{0} & \pie{360} & \pie{0} \\
\bottomrule
\end{tabular}
}
\begin{tablenotes}
\item[]~\pie{360}: knowledge is available; \pie{0}: knowledge is not available.
\end{tablenotes}
\label{tab:knowledge}
\end{table}
We detail the capabilities and knowledge of the model provider, MLaaS server, and user as follows:

\noindent  \textbf{Model Provider.} They are model developers or model owners, who have full control of the model, including the training data and white-box information of the model. After they delegate the model to MLaaS for commercialization, the control is also transferred to the server. 

\noindent  \textbf{MLaaS Server.} The server is the agency that provides MLaaS to users for normal business and also provides machine unlearning services for compliance with privacy regulations. To simulate a practical scenario, we assume the server only has a test dataset but does not have the original training dataset for two reasons. First, users' training data may contain sensitive or personally identifiable information. Thus, to ensure their privacy and confidentiality, users may not be willing to have their data stored by the server. Second, many regions and countries have strict data protection regulations that govern the collection, storage, and usage of personal data. For example, the tech giant Meta in May 2023 was fined a record 1.2 billion euros (\ie US\$1.3 billion) and ordered to stop transferring data collected from Facebook users in Europe to the United States~\cite{meta}. Thus, the server itself may choose not to store users' training data to simplify compliance with these regulations. 

\noindent  \textbf{Authorized Unlearning Users.} Authorized unlearning users are the users who are authorized to raise machine unlearning requests to the server if they wish to opt-out by submitting the unlearned data to the server. They can be the data provider for the model training. These users can also access the model provided by the server to utilize its predictive ability as normal API users. The unlearning procedure will be conducted at the server using the pre-selected unlearning strategies. 

\noindent \textbf{Machine Unlearning Threats.} 
We anticipate a scenario where some authorized users, with permission to initiate unlearning, might have malicious intent or could be compromised. These users, despite only contributing a small proportion of the training data, aim to induce severe performance degradation in the server's model by exploiting unlearning requests with a few samples. Such situations underscore the necessity for robust security measures and stringent user regulations within machine learning systems. 

\noindent \textit{Property of Malicious Unlearning.} The properties of the malicious unlearning behaviors are: 

\begin{itemize}[leftmargin=*]
\item \textit{Performance degradation.} The server's normal business performance incurs an unexpected degradation when fulfilling the user's unlearning requests: the utility of the model is heavily compromised. 
\item \textit{Stealthiness of the unlearned sample.} The manipulated unlearned sample for over-unlearning is not easy to be distinguished from normal samples. 
\item  \textit{Stealthy prediction of the unlearned sample.} Note that the authorized user could also submit a label of the unlearned sample. The submitted label of the manipulated unlearned sample may be needed to be consistent with the prediction derived from the deployed model.
\end{itemize}
The latter two properties of malicious unlearning are proposed from the MLaaS server's perspective: the server may be aware of the malicious unlearning requests and implement protections to protect its model. Specifically, the server can verify both the quality of the sample’s features and the prediction label of the unlearned sample. Thus, malicious unlearning requires to satisfy these two properties for ensuring that it can bypass basic protections implemented by the server.

\noindent \textit{Capabilities of Malicious Users.} We assume such a malicious user only has black-box access to the model of the server, which means the user can submit a data sample $\bm{x}$ to query the model and obtain a vector of probabilities $\mathcal{Y}$ but cannot know the parameters and architectures of the model. Note that black-box access assumption is very practical and also strict for the malicious user: his adversarial knowledge is similar to the knowledge of a normal user. Based on $\mathcal{D}_u$ and black-box access, the malicious user aims to construct a perturbed unlearn dataset $\mathcal{D}_{u}^{\prime}$, which will be sent to the server for unlearning as well as compromising its model. 
The practicality of this threat is also manifested in that a malicious user could theoretically superimpose an unlimited amount of information on the data set to be unlearned, causing an unanticipated decline in model performance. 
Besides, the degradation of the model's performance is multifaceted. For instance, taking the example of unlearning 100 samples from class A, malicious unlearning can easily be achieved by injecting additional information into these 100 samples. This would not only result in the over-forgetting of information from category A but may also cause the model's performance to decline across all categories. Even more, it could potentially target a specific category B by injecting information about B into these 100 samples of A, thereby accomplishing a targeted malicious unlearning.

\begin{table}[t]
\centering
\caption{An overview of threats to machine unlearning.}
\label{tab:banchmark}
\resizebox{\linewidth}{!}{

\begin{tabular}{@{}lccccc@{}}
\toprule
\multirow{3}{*}{\begin{tabular}[c]{@{}l@{}}\textbf{Threats to}\\ \textbf{Machine Unlearning}\end{tabular}} & \multicolumn{3}{c}{\textbf{Adversary's Knowledge Requirement}} & \multicolumn{2}{c}{\textbf{Unlearning Methods}}\\ 
\cmidrule(l){2-4} \cmidrule(l){5-6}  
 & \begin{tabular}[c]{@{}c@{}}Training\\ Procedure\end{tabular} & \begin{tabular}[c]{@{}c@{}}Training\\ Dataset\end{tabular} &  \begin{tabular}[c]{@{}c@{}}White-box\\ Model\end{tabular} & 
 \begin{tabular}[c]{@{}c@{}}Exact \\ unlearning\end{tabular}
 &  \begin{tabular}[c]{@{}c@{}}Approximate \\ unlearning\end{tabular}\\
\midrule
Slow-down Unlearning~\cite{marchant2022hard} & \pie{360} & \pie{360} & \pie{360} & \rect{1} & \rect{-1}\\
Camouflaged Poisoning~\cite{di2022hidden} & \pie{360} & \pie{360} & \pie{360} &  \rect{-1} & \rect{1} \\
Over-unlearning (\textbf{Ours})& \pie{0} & \pie{0} & \pie{0} & \rect{1} & \rect{-1}\\ 
\midrule
\begin{tabular}[c]{@{}l@{}}Resources Available \\ in MLaaS Scenario\end{tabular} & \pie{0} & \pie{0} & \pie{0}  & \rect{1} & \rect{-1} \\ 
\bottomrule
\end{tabular}
}
\begin{tablenotes}
\item[] \pie{360}: the knowledge is required or the resource is available; \pie{0}: the knowledge is not required or the resource is not available; \rect{-1}: the unlearning method is applicable; \rect{1}: the unlearning method is not applicable. 
\end{tablenotes}
\end{table}

\subsection{Difference from Existing Threats to Machine Unlearning} 
Currently, there are two works~\cite{marchant2022hard,di2022hidden} that investigate the potential threats to machine unlearning. The first work~\cite{marchant2022hard} proposes slow-down attacks that aim to increase the computational cost of the unlearning process by adding perturbations to the original unlearned samples. The second work~\cite{di2022hidden} proposes targeted attacks that aim to cause the model to misclassify particular target test samples. To achieve the targeted attacks in machine unlearning, the adversary first creates poison samples that contain features of the target test samples and adds them to the training dataset. Then, by submitting an unlearning request to unlearn the poison samples, the model will be triggered to make wrong predictions on the target test samples. \autoref{tab:banchmark} summarizes the two existing threats and our proposed over-unlearning threat to machine unlearning. 

Our over-unlearning threat is significantly different from the two existing threats in adversary's knowledge requirement, threat scenarios and threat goals. First, we are the first to investigate the threat under the MLaaS scenario, where the existing two threats cannot be materialized because the resources available in MLaaS are not enough for the adversary to pose the threat. Specifically, as depicted in \autoref{tab:banchmark}, the existing two threats require the adversarial to know the training procedure, have access to the training dataset, and have white-box access to the model, while our threat is more practical and does not need such adversarial knowledge. Second, the goal of over-unlearning is different from the two existing threats: over-unlearning is to compromise the model's utility, while the existing two threats focus on increasing the computational cost of unlearning or achieving targeted attacks on specific samples (\ie fool the model).

\section{Methodology of Malicious Unlearning}
In the context of machine learning, there is limited understanding of how individual data samples impact a machine learning model. Although we have the influence function~\cite{cook1980characterizations,koh2017understanding} to measure the effect of a single training sample on model parameters, it only works perfectly on convex models while cannot precisely measure the effect on complex non-convex models such as DNNs~\cite{bourtoule2021machine}. The difficulty of measuring the amount of knowledge or information contained in a single data sample provides avenues for malicious users to exploit in machine unlearning services: the user can intentionally manipulate their data so that the server unlearns the revised data having a higher effect on the model than unlearning the original data. 

In this section, we formulate the over-unlearning and introduce two possible strategies that an authorized unlearning user can leverage for malicious unlearning in MLaaS scenarios, to answer the research question ``Is it feasible for a user to compromise the normal business of the server by requesting machine unlearning services in MLaaS? And how easily can the user achieve the compromise?'' Based on the different granularity of the input space, we introduce two types of methods that the user can leverage to materialize the threats of over-unlearning. The first type works on \textit{sample-wise} modification of the data, while the second one works on \textit{pixel-wise} modification of the data.

\subsection{Problem Statement}
In this paper, we mainly focus on ML classification, as it is one of the most common ML applications. Let $(\bm{x},y)$ be a data sample with multidimensional features. An ML classifier is a function $f(\cdot)$ that takes as input $\bm{x}$ and outputs a vector of probabilities~$\mathcal{Y}$. The length of $\mathcal{Y}$ equals the number of classes in the classification task. Each entry $y_i$ in $\mathcal{Y}$ represents the posterior probability of the model assigning $\bm{x}$ to a class (\ie the label) $c_i \in \mathcal{C}$, where $\mathcal{C}$ is the set of all classes. 

\begin{figure*}[t]
\centering
\includegraphics[width=0.83\linewidth]{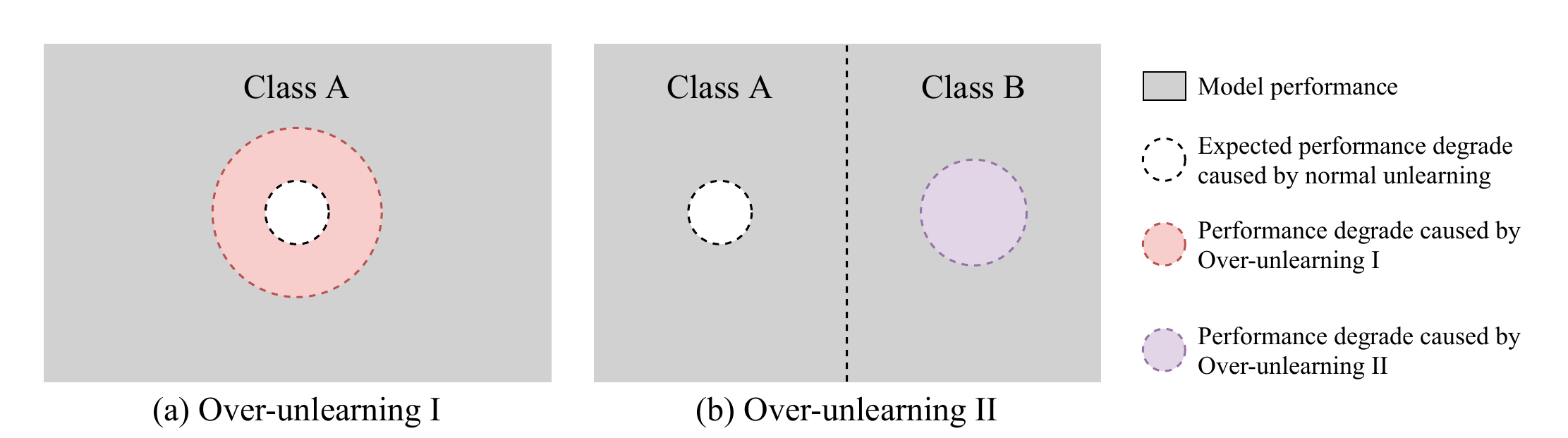}
\caption{An illustration of two types of implications of over-unlearning. The white circle represents the information that the $\bm{\theta}^*$ should unlearn.}
\label{fig:unlearning}
\end{figure*}

Let $\mathcal{M(\cdot,\cdot)}$ be a fixed unlearning method, and $\bm{\theta}^*$ be the trained model owned by the server. Let $\mathcal{D}_u \subset \mathcal{D}_{\textrm{train}}$ be an unlearned dataset owned by a user. Let $\mathcal{D}_{u}^{\prime}$ be a perturbed version of $\mathcal{D}_{u}$ generated by the user for the purpose of posing the over-unlearning threat to the server. To facilitate the understanding of over-unlearning, we first introduce the scenarios of normal unlearning and malicious unlearning.

\noindent \textbf{Normal Unlearning.} Normal unlearning is defined as the case that the user faithfully submits $\mathcal{D}_{u}$ to the server for unlearning. Based on $\mathcal{M(\cdot,\cdot)}$ and $\mathcal{D}_{u}$, the server produces an unlearned model $\bm{\theta}_{u}^*$.

\noindent \textbf{Malicious Unlearning.} Malicious unlearning is defined as the case that the user maliciously submits $\mathcal{D}_{u}^{\prime}$ to the server for unlearning. Based on $\mathcal{M(\cdot,\cdot)}$ and $\mathcal{D}_{u}^{\prime}$, the server produces an unlearned model $\bm{\theta}_{t}^*$.

Note that the server may have limited capacity to check the quality of the unlearned model, and the most feasible indicator is to check the test accuracy of the model using a test dataset. The investigation of accuracy can be conducted class by class to check the quality of each class. 
Let $\mathcal{D}_{\textrm{test}}$ be a test dataset. 
For simplicity, we consider the data $(\bm{x},y) \in \mathcal{D}_u$ hosted by the malicious unlearning user being from one specific class A. Let $\mathcal{D}_{A} \subset \mathcal{D}_{\textrm{test}}$ be a sub-test dataset containing the test samples of the class $A$ in $\mathcal{D}_{\textrm{test}}$. 
We define $\alpha_1$ as the test accuracy of $\bm{\theta}_{u}^*$ on $\mathcal{D}_{A}$ and $\alpha_2$ as the test accuracy of $\bm{\theta}_{t}^*$ on $\mathcal{D}_{A}$. 
We can define over-unlearning as follows:

\begin{definition}[Over-unlearning] We call the case Over-unlearning if the utility of $\bm{\theta}_{t}^*$ is smaller than that of $\bm{\theta}_{u}^*$ on $\mathcal{D}_{A}$, e.g., the accuracy $\alpha_2 < \alpha_1$.
    
\end{definition}
Over-unlearning represents that the information the server unlearned exceeds what the server ought to unlearn. For example, a classifier trained to determine ``dog'' and ``cat'' has a test accuracy of 90\% on the class ``dog''. Unlearning 10\% of the training data of ``dog'' produces an unlearned model with a test accuracy of 88\% on ``dog'', \ie the accuracy degradation is 2\%. We consider over-unlearning to happen if unlearning the same training data with some modifications using the same unlearning method produced an unlearned model with a test accuracy of less than 88\% on ``dog'', \eg 80\%.

Over-unlearning may result in two types of degradation, as shown in \autoref{fig:unlearning}. Over-unlearning-I highlights the degradation observed in class A when manipulating samples of class A, while Over-unlearning-II uncovers the degradation experienced in classes other than A. The impacts of these two types hinge on whether the supplementary information introduced to instigate over-unlearning regarding classes other than A or not.

\subsection{Blending as Naive Over-unlearning}
\begin{figure}[t!]
    \centering
    \subfloat[Original sample]{\includegraphics[width=0.3\linewidth]{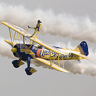}}
    \hspace{1pt}
    \subfloat[Blend sample]{\includegraphics[width=0.3\linewidth]{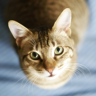}}
    \hspace{1pt}
    \subfloat[Blended sample]{\includegraphics[width=0.3\linewidth]{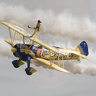}}
    \caption{An example of blending ``airplane'' sample with ``cat'' sample (with $\lambda=0.1$).}
    \label{fig::blend_demo}
\end{figure}

The most simple way to achieve over-unlearning is to incorporate additional sample information into the unlearned sample, \eg blending the original unlearned sample from class~A with a sample from another class B without introducing any computational overheads. 
Therefore, by unlearning $\mathcal{D}_{u}^*$ that contains the features of the blended sample from class B, the model will suffer an unexpected performance degradation on class B as a result of  additional information over-unlearned. 
Therefore, we first use such a lightweight sample-wise strategy to illustrate the feasibility of over-unlearning. Here, the additional information introduced to cause over-unlearning is other than the original class A, then the measured impact belongs to over-unlearning-II.
To meet the stealthiness requirement of the unlearned sample, we consider the malicious user can embed the features of the samples of the target class into the data samples in $\mathcal{D}_{u}$. 
Specifically, we consider the malicious user can use the blend technique~\cite{chen2017targeted} to embed the features of the samples of another class into the data samples in $\mathcal{D}_{u}$. The user leverages an injection function $\Pi(\cdot,\cdot)$ to blend $\bm{x}$ with $\bm{x}_b$, which is defined as follows:
\begin{equation}
    \Pi(\bm{x},\bm{x}_b)=\lambda \cdot\bm{x}+(1-\lambda)\cdot\bm{x}_b,
\end{equation}
where $\lambda \in [0,1]$ is the hyper-parameter representing the blending ratio. Here, both $\bm{x}$ and $\bm{x}_b$ are in their vector representations. The $\lambda$ can be considered as the hyperparameter to control the transparency level of the blending. We visualize an example of blending a ``airplane'' sample with a ``cat'' sample in \autoref{fig::blend_demo}.

To further satisfy the stealthiness of the manipulated unlearned sample, we modify the unlearned data $\bm{x}$ by blending it with the sample $\bm{x}_{b}$ in class B, while modifying $y$ to the predicted label of the server's model on the modified data. The submitted label is consistent with the prediction of the deployed model. From the server's perspective, it is difficult to notice the malicious behavior of the user based on the submitted label. 

\subsection{Pushing as Advanced Over-unlearning}
Even though the blending-based strategy is cost-efficient and model-irrelevant, the over-unlearning is conducted in a blind way and the degradation performance is not effective on all datasets (detailed in Section~\ref{sec:effective_blending}). Therefore, we introduce an advanced over-unlearning method. 

\noindent \textbf{Motivation.} 
The behavior of ML models is reflected in their decision boundary, which represents the region where the model assigns different class labels or makes decisions based on the input features. We have a key observation that ML models can be more confused when predicting samples that are near the decision boundary, given that even a slight change in the input can lead to different predictions by the model. Thus, we consider that samples near the decision boundary of a model are more informative than those that are farther away from it because samples near the decision boundary carry more ambiguity in their class assignments. This intuition is similar to the entropy~\cite{shannon2001mathematical} in information theory which measures the average amount of information contained in a random variable. In information theory, a random variable is considered more informative (\ie with higher entropy) when its possible outcomes are more random or unpredictable. 

\begin{figure}[t]
\centering
\includegraphics[width=0.9\linewidth]{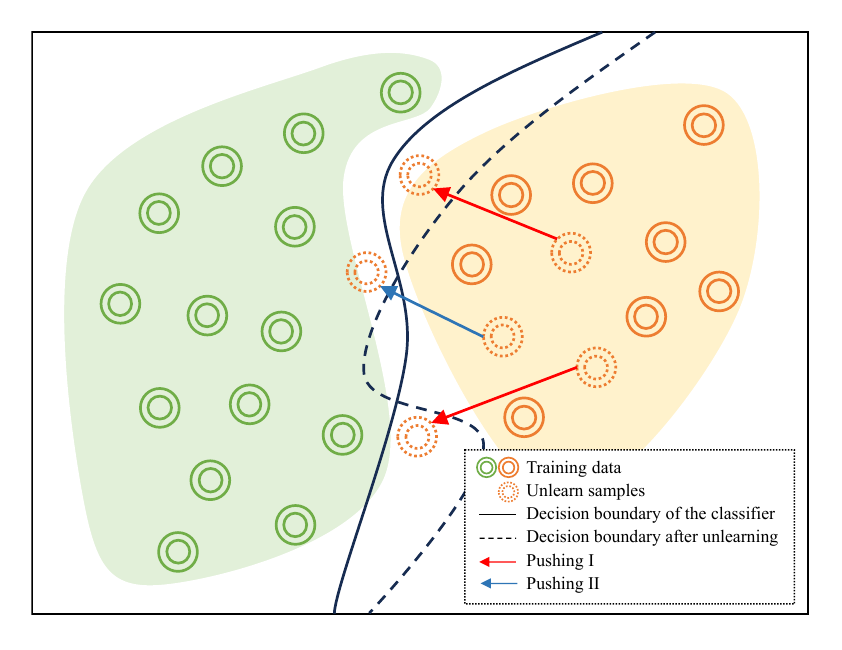}
\caption{An illustration of over-unlearning using adversarial perturbation. Moving the unlearned data to the decision boundary for unlearning can significantly change the decision boundary of the model.}
\label{fig:over-unlearning}
\end{figure}

Based on the above motivation, we consider the malicious user can enlarge the unlearning effect of the samples in $\mathcal{D}_u$ by intentionally moving them to the decision boundary of the model. 
When the data sample is moved near but still within the decision boundary, the additional information to cause over-unlearning is still about the original class~A. Therefore, the over-unlearning type is Over-unlearning-I, where the performance degradation derived from over-unlearning mainly affects class~A.
However, after the data sample is moved to cross the decision boundary, the additional information to cause over-unlearning involves classes other than class A. Correspondingly, the over-unlearning type is both Over-unlearning-I (primary) and II (secondary), where the performance degradation derived from over-unlearning affects other influenced classes.
To investigate these two scenarios, we explore two pushing-based over-unlearning methods as depicted in \autoref{fig:over-unlearning}. We describe these two methods as follows.

\noindent \textbf{Over-unlearning by Pushing-I.} This method moves the data sample toward the decision boundary but not across it. From the perspective of the model, it can still correctly predicts the label of the sample because the sample did not cross the decision boundary. Pushing-I helps to evaluate how unlearning samples near the decision boundary can affect the performance of the model.

\noindent \textbf{Over-unlearning by Pushing-II.} This method moves the data sample just across the decision boundary. After the manipulation, such data samples are beyond the prediction ability of the model because they are moved to a region other than their correct labels. Pushing-II also helps to evaluate how unlearning samples that are hard to make predictions can affect the performance of the model.

Note that samples in $\mathcal{D}_u$ themselves may be close to the decision boundary of the model. However, this does not undermine our methods: First, the majority of the training samples are not located near the decision boundary, which is often true when the model is trained on well-distributed and balanced datasets~\cite{bishop2006pattern}. Second, even if the samples in $\mathcal{D}_u$ are close to the decision boundary, we can move them further to the decision boundary. Last, we focus on investigating whether our method of moving samples to the decision boundary can enlarge their unlearning effect. As long as one of the malicious users can leverage the method to materialize the over-unlearning threat, we can consider the threat that exists in the machine unlearning services.

Let $\bm{x} \in \mathcal{D}_u$ be an unlearned sample. Based on our method, the user aims to construct a perturbed version of $\bm{x}$:
\begin{equation}
    \bm{x}'=\bm{x}+\bm{\delta},
\end{equation}
where $\bm{x}'$ is the perturbed version of $\bm{x}$, and $\bm{\delta}$ is the perturbation. $\bm{x}^\prime$ is required to satisfy:
\begin{equation}
    \textrm{Dis}(\bm{x}^\prime,\bm{\theta}^*) \leq \epsilon,
\end{equation}
where $\textrm{Dis}(\cdot,\cdot)$ is the distance of $\bm{x}^\prime$ to the decision boundary of the model $\bm{\theta}^*$, and $\epsilon$ is a small value of the distance threshold. However, calculating the exact distance of a data sample to the decision boundary of ML models, especially deep models can be challenging because of the complexity of the model and the lack of analytical solutions~\cite{samek2017explainable}. Thus, we propose to leverage adversarial perturbation~\cite{goodfellow2014explaining}, which is a practical and commonly used technique to move samples closer to the decision boundary of a model by adding small noises to them~\cite{papernot2016limitations}, while without requiring to exactly calculate $\textrm{Dis}(\bm{x}^\prime,\bm{\theta}^*)$.

Given that the user only has black-box access to the model of the server, we consider the user can leverage the adversarial technique from the black-box Carlini and Wagner (CW) adversarial attack~\cite{carlini2017towards,chen2017zoo}. 
The Carlini and Wagner (CW) attack is an optimization-based attack that aims to find the minimum amount of perturbation that when added to the input will lead to misclassification. 
The loss function for the CW attack is defined as follows:
\begin{equation}\label{equ:loss}
\begin{aligned}
\mathcal{L}(\bm{x}, \bm{x}') & = ||\bm{x} - \bm{x}'||_2^2 + c \cdot f(\bm{x}') \\
s.t. & \quad \bm{x}' \in [0, 1]^d,
\end{aligned}
\end{equation}
where $\bm{x}$ is the original input, $\bm{x}'$ is the perturbed input, $||\cdot||_2$ is the $\ell_2$-norm, and $c$ is a constant. $f(\bm{x}')$ is a function defined as:
\begin{equation}
    f(\bm{x}')=\max_{\bm{x}'} \{\max_{i \neq y_{{t}}} ([Z(\bm{x}')]_{i  })-[Z(\bm{x}')]_{y_{t}}, -k\},
\end{equation}
where $Z(\bm{x}')$ are the logits, $[Z(\bm{x}')]_i$ represents the predicted probability that $\bm{x}'$ belongs to class $i$, $y_t$ is a target class label toward misclassification,  and $k \geq 0$ is a margin parameter.

\begin{algorithm}[t]
\caption{Perturbation design for over-unlearning}
\begin{algorithmic}
\Require $\bm{x}$: original input; $T$: maximum number of iterations for the optimization algorithm
\State Initialize $\bm{x}' = \bm{x}$
\For{iteration in range($T$)}
\State Calculate loss $\mathcal{L}(\bm{x}, \bm{x}')$ \Comment{Equation \ref{equ:loss}}
\State Calculate the gradient of loss with respect to $\bm{x}'$
\State Update $\bm{x}'$ in the direction of the negative gradient
\EndFor\\
\Return $\bm{x}'$
\end{algorithmic}
\end{algorithm}

In this work, we consider the black-box setting, where the user has no knowledge about the architecture or parameters of the model. Here, the only information received from the model is the predicted class probabilities for all classes via querying input $\bm{x}$ for the deployed model. 
The CW attack can be performed using techniques such as zeroth order optimization (ZOO)~\cite{chen2017zoo}. The gradients of the model's output with respect to the input can be approximated using finite differences, \ie by querying the black-box model with slightly perturbed inputs and observing changes in the model's output. 
The loss term $f(\bm{x}')$ in ZOO can be formalized as follows:
\begin{equation}
\begin{aligned}
f(\bm{x}') & = \max_{\bm{x}'}\{\max_{i \neq y_t} (\textrm{log}[Z(\bm{x}')]_{i})-\textrm{log}[Z(\bm{x}')]_{y_t}, -k\}.
\end{aligned}
\end{equation}

Note that the $\log(\cdot)$ here is a monotonic function, lessening the dominance effect while preserving the order of confidence scores due to monotonicity. 

To perform the optimization, ZOO uses the zeroth order stochastic coordinate descent method, which can be written in the following form:
\begin{equation}
\bm{\delta}^{(t+1)}_{j} = \bm{\delta}^{(t)}_{j} - \eta \bm{g}^{(t)}_{j},
\end{equation}
where $j$ is a coordinate index of the pixel in the image $\bm{x}$, $\eta$ is the step size, and $\bm{g}^{(t)}_{j}$ is an estimate of the gradient of the loss function at iteration $t$, calculated using finite differences as follows:
\begin{equation}
\bm{g}^{(t)}_{j} = \frac{f(\bm{x} + \bm{\delta}^{(t)} + h \bm{e}_{j}) - f(\bm{x} + \bm{\delta}^{(t)})}{h}.
\end{equation}
Here, $h$ is a small constant used for the finite difference approximation, and $\bm{e}_{j}$ is the standard basis vector with the $j$-th component as 1.
Note that there are other black-box adversarial techniques such as substitute models~\cite{liu2017delving}. The zeroth order optimization is selected due to avoid the need to train a substitute model.

For a sample $(\bm{x},y_{\textrm{true}}) \in \mathcal{D}_u$, we use the adversarial technique described above to move it towards the decision boundary and obtain $(\bm{x}^{(1)},y^{(1)}),\ldots, (\bm{x}^{(t-1)},y^{(t-1)}), (\bm{x}^{(t)},y^{(t)})$. Assume $y^{(t-1)}=y_\textrm{true}$ and $y^{(t)} \neq y_\textrm{true}$, then Pushing-I method selects $\bm{x}^{(t-1)}$ as the modified sample and Pushing-II method selects $\bm{x}^{(t)}$ as the modified sample. 

\section{Experimental Settings}
In this section, we first introduce the datasets, models, and evaluation metrics used for the experiments. Then, we introduce the unlearning settings and the unlearning benchmarks used for evaluating the over-unlearning threat.
\subsection{Datasets and Models}
\noindent \textbf{Datasets.} In our experiments, we use three datasets to evaluate the proposed two methods for over-unlearning. The three datasets are benchmark datasets for image classification tasks, which cover a wide range of object categories with different learning complexities.

\begin{itemize}[leftmargin=*]
\item \textbf{CIFAR-10~\cite{krizhevsky2009learning}.} This dataset  contains 60,000 color images with 50,000 images in the training dataset and 10,000 images in the test dataset. It consists of 10 different classes, with each image labeled with one of the following classes: airplane, automobile, bird, cat, deer, dog, frog, horse, ship, or truck. Each image in the dataset has dimensions of $32\times32$ pixels.
\item \textbf{CIFAR-100~\cite{krizhevsky2009learning}.} Just like CIFAR-10, this dataset  contains 60,000 color images with 50,000 images in the training dataset and 10,000 images in the test dataset. CIFAR-100 includes 100 fine-grained classes, such as different types of animals, plants, household objects, and vehicles. The image size in the CIFAR-100 is the same as CIFAR-10 with dimensions of $32\times32$ pixels.
\item \textbf{STL-10~\cite{coates2011analysis}.}  This dataset  consists of 13,000 color images with 5,000 training images and 8,000 test images. STL-10 has 10 classes of airplanes, birds, cars, cats, deer, dogs, horses, monkeys, ships, and trucks with each image having a higher resolution of $96\times96$ pixels. Compared to the above two datasets, STL-10 can be considered as a more challenging dataset with higher learning complexity.
\end{itemize}

\noindent \textbf{Models.} We use the VGG model~\cite{simonyan2014very} and the ResNet model~\cite{he2016deep} to evaluate our proposed methods for over-unlearning. These two model architectures are benchmark classification models for image classification tasks. The detailed description of the models used in this paper is in Appendix~\ref{appendix:settigns}.

\noindent \textbf{Metric.} As the MLaaS service provider, the utility of the model is a priority of the server because it aims to offer models that are accurate for providing predictions to their users. The server may have limited capacity to check the quality of the unlearned model, and the most feasible indicator is to check the test accuracy of the model using a test dataset. We report the accuracy of the model on the test dataset to assess the utility of the model in the experiments. 

\subsection{Unlearning Settings and Benchmarks}
\noindent \textbf{Number of Unlearned Samples.} To mimic a practical scenario where the malicious user only contributes a small proportion of training data, we assume the user has no more than 50\% training data of a class that can be modified for over-unlearning. Specifically, we assume the user has no more than 200, 200, and 2,000 samples in CIFAR-100, STL-10, and CIFAR-10, respectively.

\noindent \textbf{Perturbation Magnitude for Pushing.} To ensure the stealthy of the modified samples in pushing methods, the perturbation added to the original sample should be imperceptible. In our experiments, we bound the perturbations by $\ell_2$-norm perturbations of 20 to ensure the stealthiness of the modified samples. We provide visualization of modified samples with the maximum perturbation and analysis of the perceptual similarity between the modified samples and the original samples in Appendix~\ref{appendix:similarity} and \autoref{tab:ab_decp_blend}. The modified samples have high perceptual similarity with the corresponding original samples (SSIM~\cite{wang2004image} mean value around $0.97$, the closer to 1, the more similar, LPIPS~\cite{zhang2018unreasonable} mean value around $0.03$, the smaller, the more similar). 

\noindent \textbf{Unlearning Benchmarks.} We evaluate the effectiveness of our proposed methods for over-unlearning on two benchmark unlearning methods: one is a state-of-the-art gradient-based unlearning method~\cite{warnecke2023machine}, and the other is an empirical unlearning method of fine-tuning~\cite{xu2023machine} that has been widely used as the unlearning baseline. Both of the two unlearning methods are feasible for machine unlearning services where the server requires to perform unlearning with the usage of only the unlearned data sent from the users. 

\begin{itemize}[leftmargin=*]
\item \textbf{Fine-tuning based Unlearning Method.} Fine-tuning has been widely used as an empirical unlearning baseline in existing works~\cite{golatkar2020eternal,golatkar2020forgetting,graves2021amnesiac,xu2023machine}. This method aims to let the model output wrong predictions on the unlearned samples by fine-tuning the model on the unlearned samples with wrong labels. As an empirical unlearning method, the intuition of fine-tuning is the privacy information of the unlearned sample cannot be inferred because the model outputs wrong predictions. However, information of the unlearned samples might still be inferred from the parameters of the model. In our experiments, we follow the same strategy that the server randomly relabels the images sent from the user and fine-tunes the model on relabeled images for unlearning.
\item \textbf{Gradient-based Unlearning Method.} We evaluate on a state-of-the-art gradient-based unlearning method developed by Warnecke~\etal~\cite{warnecke2023machine}, which is detailed introduced by \autoref{equ:un1} and \autoref{equ:un2} in Appendix~\ref{appendix:unlearningeq}. The intuition is to leverage an irrelevant sample to overwrite the unlearned sample. In our experiments, we adopt the hyper-parameter settings in \cite{warnecke2023machine} and use the images consisting of random noise as irrelevant samples to overwrite the unlearned images. 
\end{itemize}

Compared to the fine-tuning based unlearning method, the state-of-the-art gradient-based unlearning method guarantees that the unlearned model approximates the model that is retrained on the training dataset with the unlearned data removed, \ie the removal in the unlearned model produced by the state-of-the-art gradient-based unlearning method is certified. Thus, in the experiments, we mainly report the results on the gradient-based unlearning method, while we provide the experimental results showing the effectiveness of our methods on fine-tuning in Appendix~\ref{appendix:fine}.

Note that our proposed methods may also allow a malicious user to achieve over-unlearning in MLaaS via machine unlearning requests when the server uses other unlearning methods. In our experiments, we aim to show that over-unlearning can be materialized by our proposed methods when the server uses the state-of-the-art gradient-based unlearning method or the widely used empirical unlearning method. The investigation of achieving over-unlearning by new techniques or the effectiveness of our methods on new unlearning methods is orthogonal to our work.

\section{Evaluation}
In this section, we first demonstrate the performance of the naive over-unlearning method of blending to demonstrate the feasibility of over-unlearning. Then, we present the effectiveness of the advanced over-unlearning methods of Pushing-I and Pushing-II for achieving over-unlearning\footnote{The source code of the experiments is available at \url{https://github.com/TASI-LAB/Over-unlearning}}. 

\subsection{Effectiveness of Blending Method}\label{sec:effective_blending}
To demonstrate the effectiveness of the blending method for over-unlearning, we conduct experiments on CIFAR-10, CIFAR-100, and STL-10 using the VGG model. we consider the malicious user has no more than 50\% of the training samples of ``airplane'', ``apple'', and ``airplane'' in CIFAR-10, CIFAR-100, and STL-10, respectively. We select the class of ``cat'' for CIFAR-10, ``orange'' for CIFAR-100, and ``cat'' for STL-10 to embed the information of another class into the unlearned samples of the user. Here, both
the class of the unlearned data and the additional class are randomly selected. Because the additional information is from the additional class, we mainly report the test accuracy of the model on the additional class, while reporting the overall test accuracy on all classes in \autoref{tab:acc_overall_blend} in the Appendix.

\begin{table}[t]
\centering
\caption{Effectiveness of the blending method for over-unlearning-II when unlearning 10\% and 50\% training data of a class on CIFAR-10, CIFAR-100, and STL-10.}
\label{tab:decp_effective}
\resizebox{0.78\linewidth}{!}{%
\begin{tabular}{lccc}
\toprule
\textbf{Dataset} & \begin{tabular}[t]{@{}c@{}}\textbf{\# of}\\\textbf{Samples}\end{tabular} & \begin{tabular}[t]{@{}c@{}}\textbf{Blending}\\\textbf{Ratio}\end{tabular} & \begin{tabular}[t]{@{}c@{}}\textbf{Performance}\\\textbf{Degradation}\end{tabular} \\
\midrule
\multirow{4}{*}{CIFAR-10} &\multirow{2}{*}{400} & 0.3  & 0.7\%  $\downarrow$ \\
 & &  0.5  & 1.4\% $\downarrow$  \\
\cmidrule(l){2-4}

& \multirow{2}{*}{2,000} & 0.3  & 2.8\% $\downarrow$  \\
 & & 0.5  &  8.1\% $\downarrow$\\
\midrule

\multirow{4}{*}{CIFAR-100} &\multirow{2}{*}{40} & 0.3  &\pie{0} \\
 &  & 0.5  &\pie{0} \\
\cmidrule(l){2-4}

& \multirow{2}{*}{200}  & 0.3  & \pie{0} \\
 &  & 0.5  & \pie{0}\\
\midrule

\multirow{4}{*}{STL-10} &\multirow{2}{*}{40} & 0.3  & \pie{0}\\
 &  & 0.5  & \pie{0} \\
\cmidrule(l){2-4}

& \multirow{2}{*}{200}  & 0.3  & \pie{0} \\
 &  & 0.5  & \pie{0} \\

\bottomrule
\end{tabular}
}
\begin{tablenotes}
\item[] \pie{0}: We did not observe performance degradation compared to normal unlearning.
\end{tablenotes}
\end{table}

As depicted in \autoref{tab:decp_effective}, we can see that the blending method is effective on CIFAR-10, which demonstrates the feasibility of over-unlearning. When unlearning 400 samples of ``airplane'' containing information of ``cat'' on CIFAR-10, the blending method can degrade around 1.4\% accuracy of the unlearned model on ``cat'' compared to normal unlearning. More performance degradation of 8.1\% can be achieved when the number of unlearned samples increased to 2,000. However, we did not observe the effectiveness of the blending method on CIFAR-100 and STL-10. This indicates that the naive over-unlearning method of simply blending features of the samples in another class cannot achieve over-unlearning on complex datasets, which highlights the necessity of advanced methods for over-unlearning.

\begin{mdframed}[backgroundcolor=white!10,rightline=true,leftline=true,topline=true,bottomline=true,roundcorner=2mm,everyline=true]
\textbf{Takeaway 1~}
The naive method of blending for over-unlearning is only effective in the simple classification task, while it cannot be applied to complex tasks with many class categories or complex patterns.
\end{mdframed}

\subsection{Effectiveness of Pushing-I and Pushing-II}\label{sec:exp_eff1}
To demonstrate the effectiveness of Pushing-I and Pushing-II for over-unlearning, we conduct experiments on CIFAR-10, CIFAR-100, and STL-10 using the VGG model. We consider the malicious user has no more than 50\% of the training samples of ``airplane'', ``apple'', and ``airplane'' in CIFAR-10, CIFAR-100, and STL-10, respectively. Note that the class of the unlearned data is randomly selected and we will demonstrate the choice of the class does not affect the effectiveness of our methods in the ablation study. For Pushing-I, we move the data sample near the decision boundary. For Pushing-II, we move the data sample across the decision boundary, \ie move the unlearned data to a decision region other than its original label. 

\begin{table*}[t!]
    \centering
    \caption{The overall accuracy of the model before unlearning, with normal unlearning, with Pushing-I, and with Pushing-II on CIFAR-10, CIFAR-100, and STL-10.}
    \resizebox{1\textwidth}{!}{%
    \begin{tabular}{lcccccccc}
    \toprule
         &  & & \multicolumn{3}{c}{Test accuracy when unlearning 10\% data of a class} & \multicolumn{3}{c}{Test accuracy when unlearning 50\% data of a class} \\
       \cmidrule(lr){4-6}  \cmidrule(l){7-9}
        Dataset & Training accuracy & Test accuracy & Normal unlearning & Pushing-I & Pushing-II & Normal unlearning & Pushing-I & Pushing-II  \\
       \midrule
       CIFAR-10 &  81.5\% & 79.8\% & 79.3\% & 73.5\% (5.8\%$\downarrow$) & 73.7\% (5.6\%$\downarrow$) & 78.7\% & 66.4\% (12.3\%$\downarrow$) & 66.9\% (11.8\%$\downarrow$)\\
       CIFAR-100 & 76.3\% & 51.1\% & 51.1\% & 50.7\% (0.4\%$\downarrow$) & 50.7\% (0.4\%$\downarrow$) & 49.7\% & 49.1\% (0.6\%$\downarrow$) & 49.2\% (0.5\%$\downarrow$) \\
       STL-10 & 96.3\%& 56.6\% & 56.6\% & 56.2\% (0.4\%$\downarrow$) & 49.7\% (6.9\%$\downarrow$) & 50.2\% &  49.5\% (0.7\%$\downarrow$) & 32.8\% (17.4\%$\downarrow$) \\
       \bottomrule
    \end{tabular}
}
    \label{tab:acc_pushing_overall}
\end{table*}

\begin{figure}[t!]
    \centering
  \subfloat[Unlearn 10\% data of a class]{%
       \includegraphics[width=0.24\textwidth]{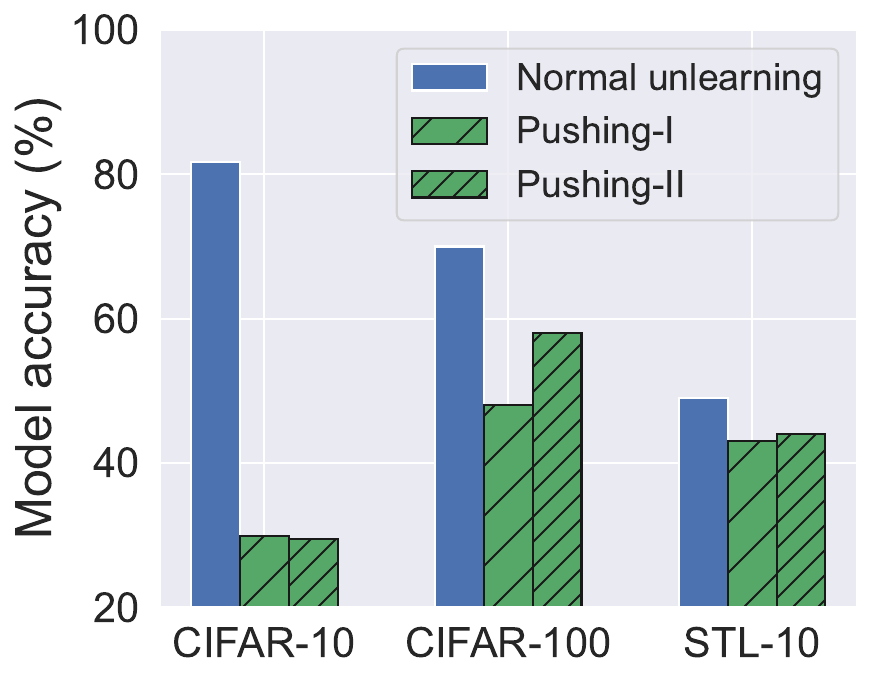}}
  \subfloat[Unlearn 50\% data of a class]{%
        \includegraphics[width=0.24\textwidth]{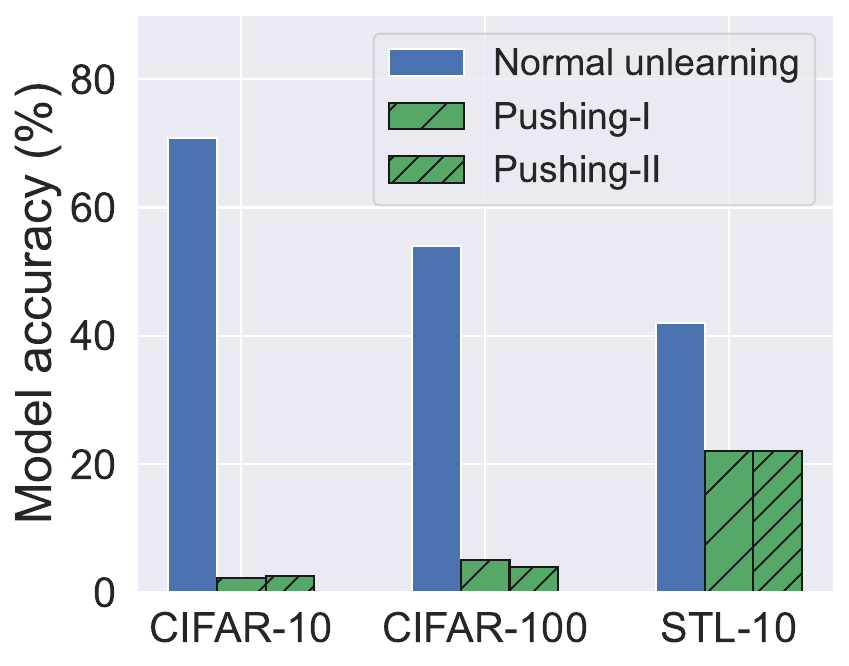}}
  \caption{Effectiveness of Pushing-I and Pushing-II for over-unlearning-I when unlearning 10\% and 50\% training data of a class on CIFAR-10, CIFAR-100, and STL-10.}
  \label{fig:overun}
\end{figure}

As we can see in \autoref{tab:acc_pushing_overall}, both Pushing-I and Pushing-II can degrade the overall accuracy of the model compared to normal unlearning, which indicates the effectiveness of the two methods for Over-unlearning-I and Over-unlearning-II. To better demonstrate how severe over-unlearning can be achieved by Pushing-I and Pushing-II, we report the test accuracy of the model on the class that the unlearned data comes from, \ie reporting the performance for Over-unlearning-I in the following parts.

\begin{figure}[t!]
    \centering
    \subfloat[Original sample]{\includegraphics[width=1.in]{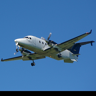}}
    \hspace{2pt}
    \subfloat[Pushing-I]{\includegraphics[width=1.in]{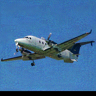}}
    \hspace{2pt}
    \subfloat[Pushing-II]{\includegraphics[width=1.in]{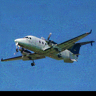}} 
    \\
    \subfloat[Original sample]{\includegraphics[width=1.in]{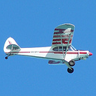}}
    \hspace{2pt}
    \subfloat[Pushing-I]{\includegraphics[width=1.in]{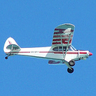}}
    \hspace{2pt}
    \subfloat[Pushing-II]{\includegraphics[width=1.in]{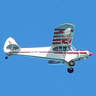}}
    \caption{Stealthiness of the modified examples in the Pushing method: STL-10 samples of ``airplane'' used for Pushing-I and Pushing-II. Human inspection cannot easily notice the existence of perturbations in the unlearned data.}    
    \label{fig::decep_demo}
\end{figure}

As depicted in \autoref{fig:overun}, we can see that both Pushing-I and Pushing-II can achieve the goal of Over-unlearning-I on all datasets: the utility of the model is smaller than the model under normal unlearning. For example, as depicted in \autoref{fig:overun}(a), when unlearning 10\% training data of a class in CIFAR-10, Pushing-I and Pushing-II can reduce the utility of the model from 87.8\% to around 29.5\%, while normal unlearning has the accuracy of the model of 81.7\%. Over-unlearning becomes more severe when unlearning 50\% training data of the class. As depicted in \autoref{fig:overun}(b), Pushing-I and Pushing-II can reduce the utility of the model on that class to around 2\%, indicating that the model becomes useless on predicting the class of the unlearned data. The results in \autoref{fig:overun} demonstrate the effectiveness of Pushing-I and Pushing-II in achieving the property of performance degradation in malicious unlearning. To demonstrate the property of the stealthiness of the unlearned samples in Pushing-I and Pushing-II, we visualize two ``airplane'' samples in STL-10 in  \autoref{fig::decep_demo}. As we can see, human inspection cannot easily notice the existence of perturbations in the unlearned data.

\begin{figure}[t!]
    \centering
  \subfloat[Baseline and Push-I]{%
       \includegraphics[width=0.24\textwidth]{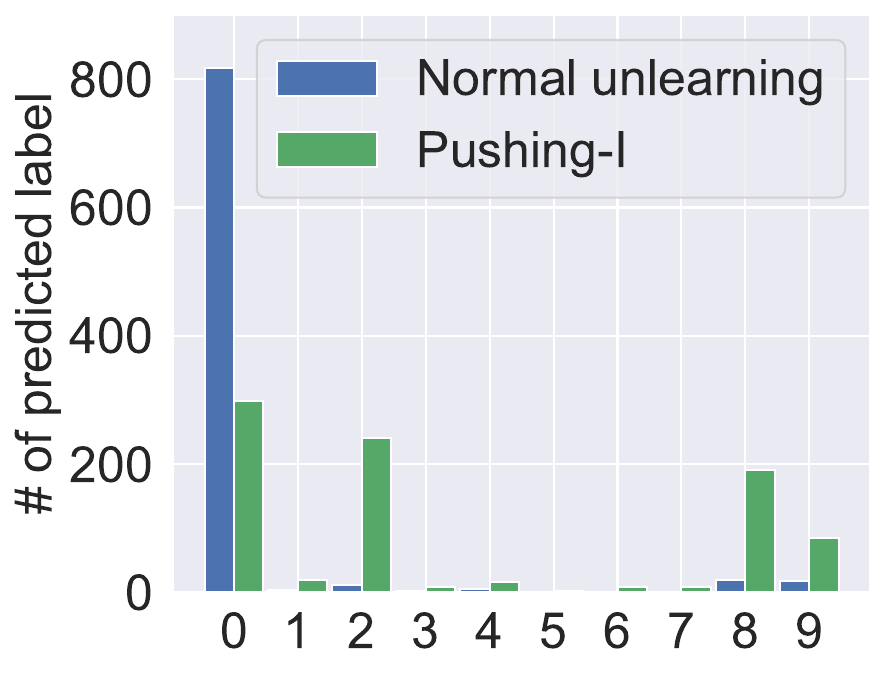}}
  \subfloat[Baseline and Push-II]{%
        \includegraphics[width=0.24\textwidth]{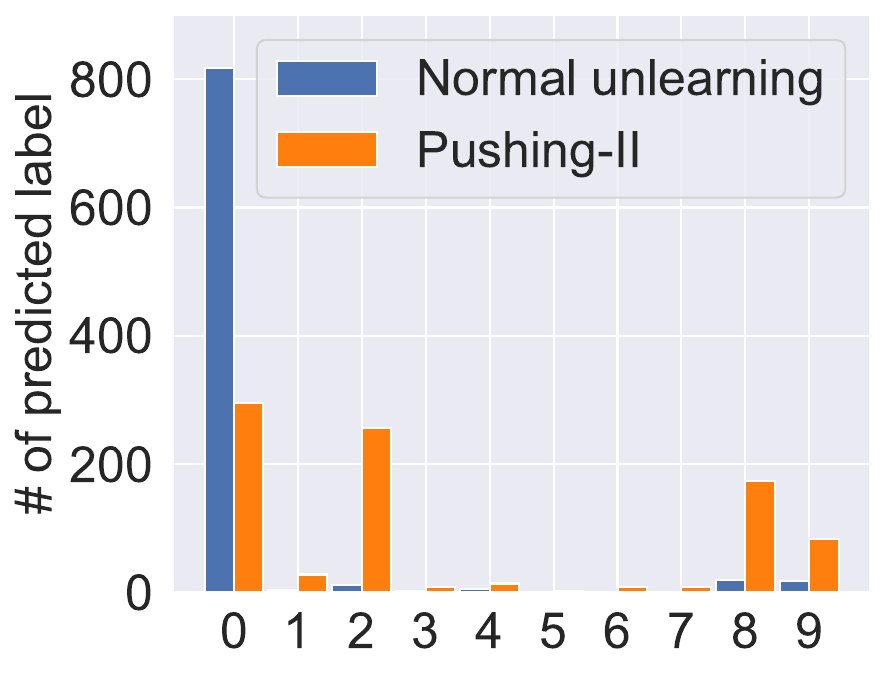}}
  \caption{Prediction distribution comparison between the model produced by normal unlearning and the models produced by Pushing-I and Pushing-II when unlearning 10\% samples of ``airplane'' on CIFAR-10. The x-axis represents the category that exists in the datasets and ``0'' represents the class ``airplane'' of the unlearned data. The y-axis represents the number of predicted labels under that category. }
  \label{fig:overun_distribution}
\end{figure}

\begin{figure}[t!]
    \centering
  \subfloat[Baseline and Push-I]{%
       \includegraphics[width=0.24\textwidth]{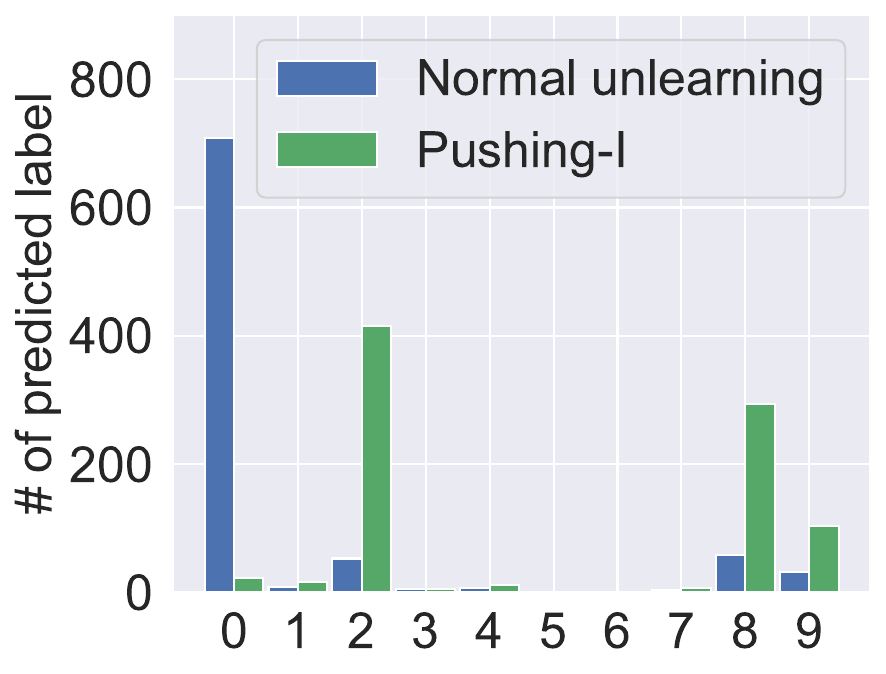}}
  \subfloat[Baseline and Push-II]{%
        \includegraphics[width=0.24\textwidth]{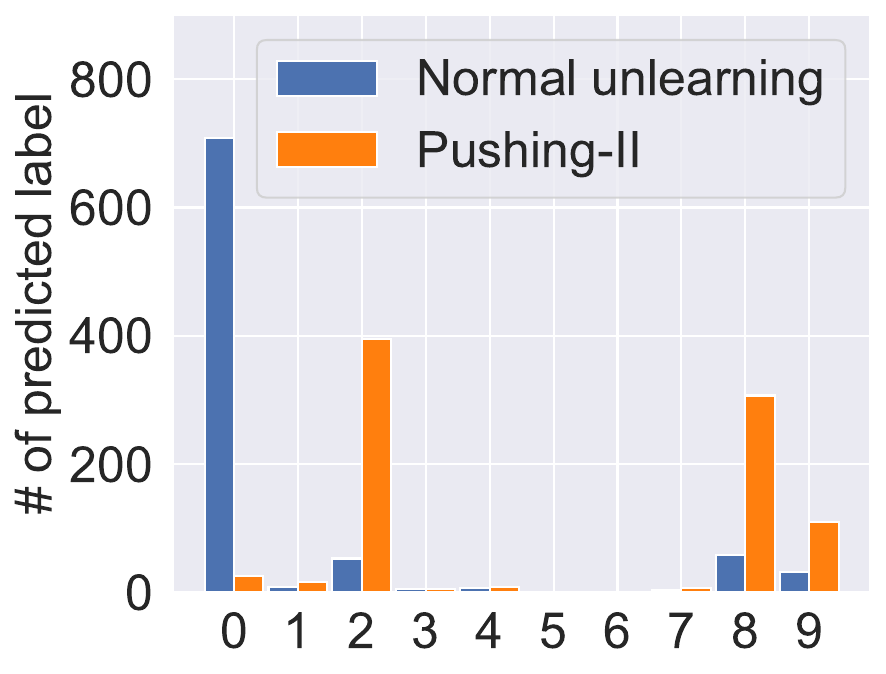}}
  \caption{Prediction distribution comparison between the model produced by normal unlearning and the models produced by Pushing-I and Pushing-II when unlearning 50\% samples of ``airplane'' on CIFAR-10.}
  \label{fig:overun_distribution_append}
\end{figure}

To study how the unlearned model produced by malicious unlearning is different from the unlearned model produced by normal unlearning, we visualize the prediction distribution of the unlearned model on CIFAR-10 in \autoref{fig:overun_distribution}. As we can see, the prediction distribution of the baseline model produced by normal unlearning is very different from that of the unlearned model produced by Pushing-I or Pushing-II. After normal unlearning 400 training samples of ``airplane'', there are 817 test samples of ``airplane'' that can be corrected and predicted as ``airplane'' by the unlearned model. However, when malicious unlearning of Pushing-I or Pushing-II happens, the over-unlearned model can only predict correctly on around 290 test samples, while largely assigning the rest of the testing samples to the label of ``2'' (``bird'') and ``8'' (``ship''). This phenomenon exacerbates when unlearning 2,000 training data of that class as depicted in \autoref{fig:overun_distribution_append}.

Motivated by \autoref{fig:overun_distribution} and \autoref{fig:overun_distribution_append} that most of the wrongly predicted labels are assigned to ``bird'' or ``ship'', we further ask that is it possible to control the wrongly predicted label of the unlearned model on test samples of ``airplane'' in Over-unlearning-I. If so, Over-unlearning-I can be more dangerous than just reducing the utility of the model. To investigate this possibility, we select a particular decision region to move the unlearned data, \ie we move all the unlearned data to the particular decision region.

We conduct an experiment on CIFAR-10 to investigate whether it is possible to control the wrongly predicted label of the unlearned model. We leverage Pushing-I and Pushing-II methods to move the unlearned samples of ``airplane'' near or across the decision region of ``cat''. \autoref{tab:targeted_cifar10} shows the prediction results of the predicted labels of the model on CIFRA-10. As we can see, before unlearning, there are 878 test samples of ``airplane'' that are correctly predicted by the trained model. When unlearning 400 samples under normal unlearning, the model can predict most of the testing samples as ``airplane'' and predict only two test samples of ``airplane'' as ``cat''. However, under malicious unlearning, the model increases the wrong predictions on ``cat'' to 20 (Pushing-I) and 25 (Pushing-II). The wrong predictions on ``cat'' increase greatly when the user can modify 2,000 samples: there are more than 245 (Pushing-II) and 375 (Pushing-I) wrong predictions of the unlearned model on ``cat''. This demonstrates that the wrongly predicted label of the unlearned model can indeed be controlled by the malicious user by moving the unlearned data to a fixed decision region of a class. The results in \autoref{tab:targeted_cifar10} demonstrate the pushing method can be more dangerous than just reducing the utility of the model. We find the same conclusion on CIFAR-100, and the results are demonstrated in  \autoref{tab:targeted_cifar100}.

\begin{table*}[t!]
    \centering
    \caption{The predicted labels of the model on CIFAR-10 before unlearning, with normal unlearning, and with malicious unlearning. ``airplane'' is the label of the unlearned data and ``cat'' is the target label.}
    \resizebox{0.95\linewidth}{!}{%
    \setlength{\aboverulesep}{0pt}
    \setlength{\belowrulesep}{0pt}
    \setlength{\extrarowheight}{.75ex}
    \begin{tabular}{cl>{\columncolor[gray]{0.85}}ccc>{\columncolor[gray]{0.85}}cccccccc}
    \toprule
        Number of &  & \multicolumn{10}{c}{Number of the different predicted labels of the model} \\
       \cmidrule{3-12}
       unlearned samples & Model status &  Airplane & Automobile & Bird & Cat & Deer& Dog & Frog & Horse & Ship & Truck \\
       \midrule
      0 & Before unlearning & 878 & 0 & 0 & 0 & 0 & 0 & 0 & 0 & 0 & 0 \\
      \cmidrule{2-12}
      \multirow{3}{*}{400} & Normal unlearning & 817 &	4& 11&	2&	5&	1&	1&	0&	19&	18 \\
      & Pushing-I & 509 &	18&	128&	25&	16&	1&	11&	5&	127&	38  \\
      & Pushing-II &538&	16&	117&	20&	13&	1&	11&	5&	117&	40 \\
      \cmidrule{2-12}
      \multirow{3}{*}{2,000} & Normal unlearning & 708	&9&	53&	5&	7&	1&	1&	4&	58&	32\\
      & Pushing-I & 26	&34&	100&	378&	20&	1&	2&	15&	214&	88 \\
      & Pushing-II &31	&35&	164&	247&	19&	1&	3&	14&	281&	83 \\
       \bottomrule
    \end{tabular}
    }
    
    \label{tab:targeted_cifar10}
\end{table*}

\begin{table}[t]
    \centering
    \caption{The predicted labels (Top-5) of the model on CIFAR-100 before unlearning, with normal unlearning, and with malicious unlearning. ``apple'' is the label of the unlearned data. ``sweet pepper'' is the target label.}
    \resizebox{\linewidth}{!}{%
    \setlength{\aboverulesep}{0pt}
    \setlength{\belowrulesep}{0pt}
    \setlength{\extrarowheight}{.75ex}
    \begin{tabular}{cl>{\columncolor[gray]{0.85}}cc>{\columncolor[gray]{0.85}}cccc}
    \toprule
        Number of &  & \multicolumn{5}{c}{Number of the predicted labels of the model} \\
       \cmidrule(l){3-7}
        samples & Model status &  apple & pear & sweet pepper & orange & trout \\
       \midrule
      0 & Before unlearning & 71 & 0 & 0 & 0 & 0 \\
      \cmidrule(l){2-7}
      \multirow{3}{*}{40} & Normal unlearning & 70 &	0& 1&	0&	0 \\
      & Pushing-I & 57 &	2& 10&	2&	0 \\
      & Pushing-II &59&	3&	8&	1&	0 \\
      \cmidrule(l){2-7}
      \multirow{3}{*}{200} & Normal unlearning & 54	&4&	13&	0&	0 \\
      & Pushing-I & 2	&7 & 	41 & 	19&	2 \\
      & Pushing-II &5	&7&	45&	14&	0 \\
       \bottomrule
    \end{tabular}
    }
    
    \label{tab:targeted_cifar100}
\end{table}

\begin{mdframed}[backgroundcolor=white!10,rightline=true,leftline=true,topline=true,bottomline=true,roundcorner=2mm,everyline=true]
\textbf{Takeaway 2~}
\begin{itemize}
\item The proposed Pushing-I and Pushing-II methods are effective and reliable in achieving over-unlearning.
\item The wrongly predicted label of the unlearned model can be controlled in over-unlearning, which is a more severe threat than just reducing the utility of the model.
\end{itemize}
\end{mdframed}
\section{Ablation Study}

\subsection{Ablation Study for the Blending Method}
Although the blending method cannot achieve over-unlearning in complex classification tasks, it works at a certain level on simple tasks with the advantage of no computational overheads. Thus, for simple classification tasks, malicious users may still consider the blending method as an option for over-unlearning. We study how the number of unlearned samples and the blending ratio may affect the performance of the blending method. We also study whether the blending method is generic when the embedded information is from different classes than the class we have evaluated.

\noindent \textbf{Number of Unlearned Samples.} We first study how the number of unlearned samples can affect the performance of the blending method. Intuitively, the more data the malicious user has, the more information of the additional class the user can embed, which should cause a larger effect on the unlearned model on that class. We set the blending ratio to 0.3. We use the model of VGG and vary the number of unlearned samples from 400, 800, 1,200, 1,600, to 2,000. We embed ``cat'' into the unlearned data of ``airplane'' in CIFAR-10. We report the difference (\ie performance degradation) between the accuracy on the ``cat'' class of the model under normal unlearning and the model produced by the blending method.

\begin{figure}[t!]
    \centering
  \subfloat[Vary \# of unlearned samples]{%
       \includegraphics[width=0.24\textwidth]{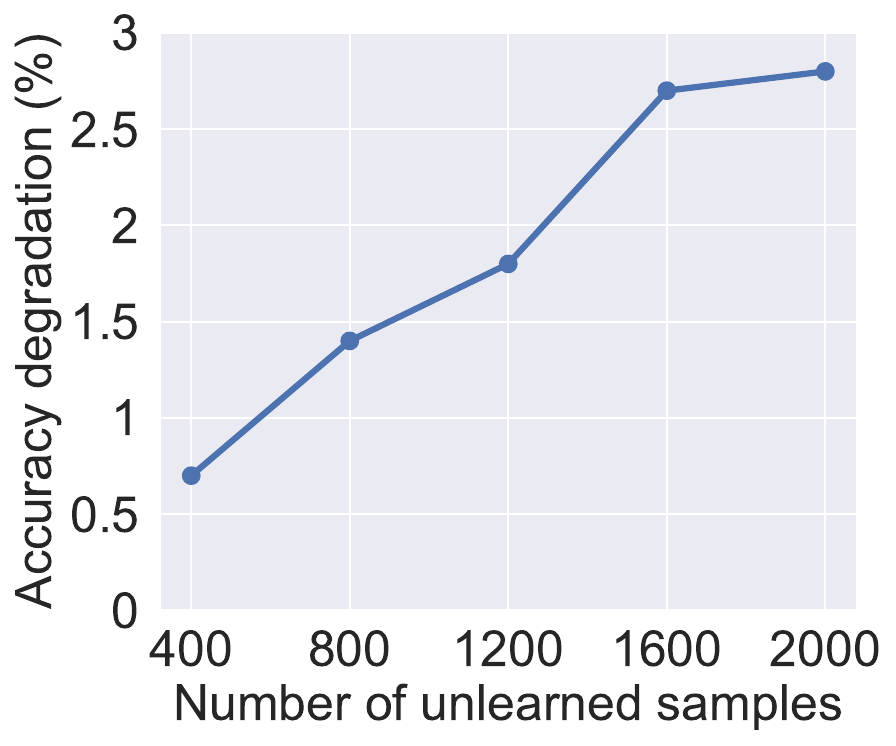}}
  \subfloat[Vary the blending ratio]{%
        \includegraphics[width=0.24\textwidth]{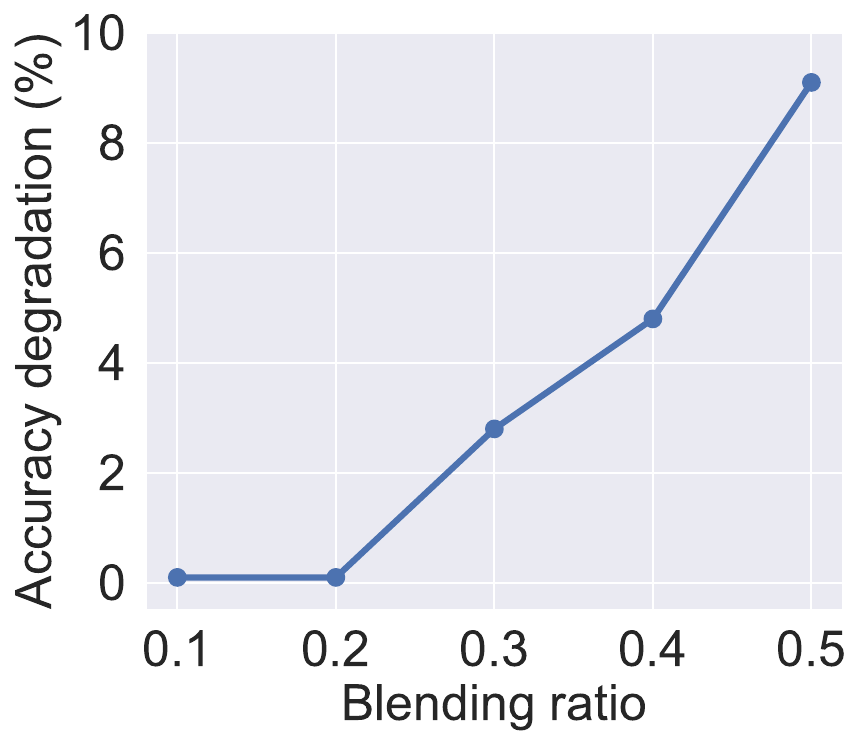}}
  \caption{Accuracy degradation of the model w.r.t. the number of unlearned samples and the blending ratio in the blending method. In general, more unlearned samples and a higher blending ratio can result in higher accuracy degradation of the model.}
  \label{fig:ab_blend}
\end{figure}

As we can see in \autoref{fig:ab_blend}(a), a larger number of unlearned samples can enable the blending method to be more effective, which aligns with our intuition. When the malicious user has 400 samples, the blending method can achieve an accuracy degradation of around 0.7\% on the model compared to normal unlearning. When having 2,000 unlearned samples, the blending method can achieve around 2.8\% accuracy degradation. Although the performance degradation caused by the blending method is not large, this naive approach demonstrates the feasibility of over-unlearning.

\noindent \textbf{Blending Ratio.} We study how the blending ratio affects the effectiveness of the blending method. Intuitively, the larger the blending ratio is, the more information of the additional class can be embedded into the unlearned data. We use the VGG model and set the number of unlearned samples to 400. We vary the blending ratio from 0.1, 0.2, 0.3, 0.4, to 0.5 and embed ``cat'' into the unlearned data of ``airplane'' in CIFAR-10.

\autoref{fig:ab_blend}(b) shows the accuracy of the unlearned model produced by the blending method under different blending ratios. As we can see, the blending method is more effective when setting a higher blending ratio. However, with a higher blending ratio, the pattern of the sample in the other class is more obvious, which can make the injected information of the other class to be less stealthy.

\noindent \textbf{Class Options.} We study whether the blending method works on other classes than the class we have evaluated. We use the VGG model and set the number of unlearned samples to 2,000 and the blending ratio to 0.3. We embed the information of the classes of ``bird'', ``horse'', and ``ship'' respectively, into the samples of ``airplane'' in CIFAR-10.  

\autoref{tab:ab_decp_class} shows the accuracy of the unlearned model under normal unlearning and the blending method on ``bird'', ``horse'', and ``ship''. As we can see, the blending method only reduces the accuracy of the model slightly compared to normal unlearning. This suggests that the naive blending for over-unlearning method may not be generic and reliable, which highlights the importance of using the advanced pushing method to achieve over-unlearning.

\begin{table}[t]
    \centering
    \caption{Unlearned model accuracy w.r.t. different classes under the blending method.}
    \resizebox{\linewidth}{!}{%
    \begin{tabular}{lccc}
    \toprule
         & Bird & Horse & Ship \\
         \midrule
        Normal unlearning & 74.6\% & 80.5\% & 94.0\% \\
        Blending & 74.4\% (0.2\%$\downarrow$) & 80.4\% (0.1\%$\downarrow$) & 93.8\% (0.2\%$\downarrow$) \\
        \bottomrule
    \end{tabular}
    }
    
    \label{tab:ab_decp_class}
\end{table}

\begin{mdframed}[backgroundcolor=white!10,rightline=true,leftline=true,topline=true,bottomline=true,roundcorner=2mm,everyline=true] 
\textbf{Takeaway 3~} 
With a greater number of unlearned samples and higher blending ratios, the naive blending method can achieve over-unlearning more effectively.
\end{mdframed}

\subsection{Ablation Study for Pushing-I and Pushing-II}
\noindent \textbf{Number of Unlearned Samples.} We first study how the number of unlearned data can affect the performance of Pushing-I and Pushing-II. Intuitively, the more data the malicious user has, the more effect on the model the user can cause. We use the VGG model and vary the number of unlearned data samples from 400, 800, 1,200, 1,600 to 2,000 in the class ``airplane'' in CIFAR-10. We report the difference (\ie performance degradation) between the accuracy of the model produced by normal unlearning and the model produced by Pushing-I and Pushing-II.

\begin{figure}
    \centering
    \includegraphics[width=0.35\textwidth]{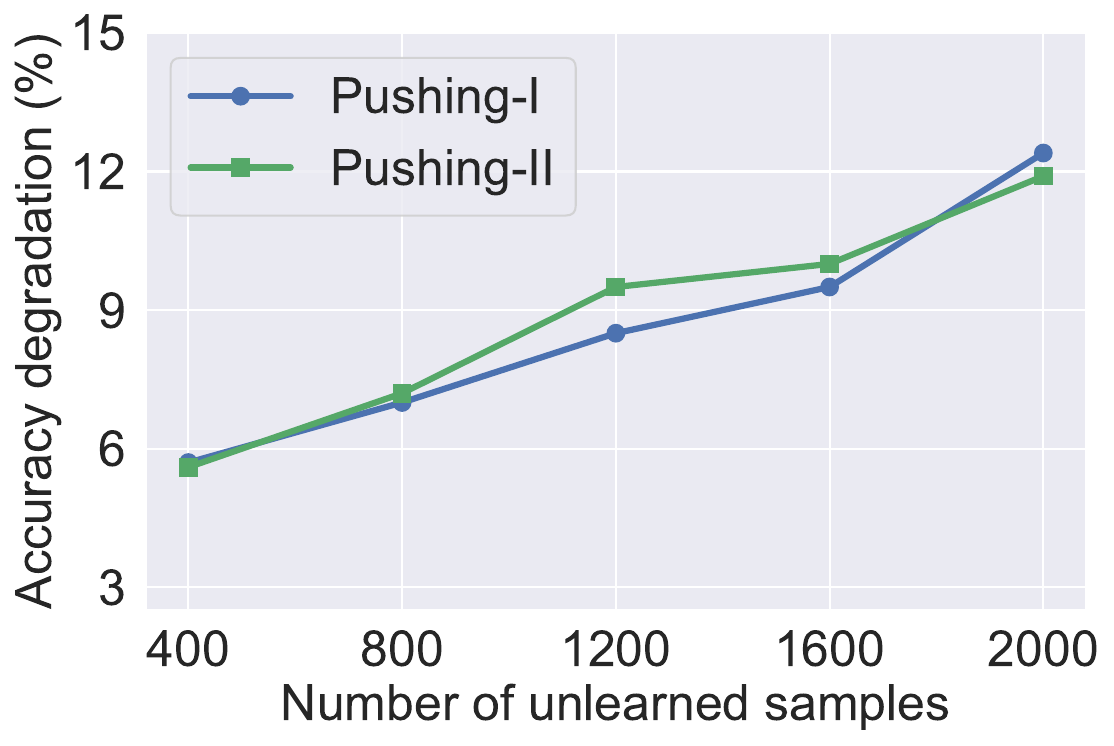}
    \caption{Accuracy degradation of the model w.r.t. number of unlearned samples. In general, more unlearned samples can result in higher accuracy degradation of the model.}
    \label{fig:ab_overI_number}
\end{figure}

As we can see in  \autoref{fig:ab_overI_number}, a larger number of unlearned samples can enable both Pushing-I and Pushing-II to be more effective. When the malicious user has 400 samples, Pushing-I and Pushing-II can degrade around 6\% accuracy of the model compared to normal unlearning. When having 2,000 unlearned samples, Pushing-I and Pushing-II can achieve around 12\% accuracy degradation, which is twice compared to the case of having 400 samples. Comparing Pushing-I and Pushing-II, it seems each has respective advantages when the malicious user has a different number of samples. However, note that Pushing-I has the advantage of sample stealthiness of both the feature and the label. Thus, Pushing-I may be more preferred by the malicious user for achieving over-unlearning.

\noindent \textbf{Model Architecture.} We study how different model architectures may affect the effectiveness of Pushing-I and Pushing-II. We use CIFAR-10 and set the number of unlearned samples in the class ``airplane '' to 400 in CIFAR-10. We use two different model architectures of VGG and ResNet. 

\autoref{tab:ab_over_model} demonstrates the effectiveness of Pushing-I and Pushing-II across VGG and ResNet. As we can see, Pushing-I can degrade the model accuracy to around 5.7\% and 2.5\% on VGG and ResNet compared to normal unlearning, and Pushing-II can degrade the model accuracy to around 5.6\% and 1.1\%. The effectiveness of Pushing-I and Pushing-II on both model architectures implies that our proposed strategy of moving data to the decision boundary of the model for achieving over-unlearning is model-agnostic. Comparing the accuracy degradation across VGG and ResNet, we find that VGG suffers more accuracy degradation than ResNet, while VGG has higher accuracy than ResNet when the data is normally unlearned. This suggests that models with higher utility might be more vulnerable to the threat of over-unlearning, which further implies the importance of the investigation of malicious unlearning.

\begin{table}[t!]
    \centering
    \caption{Unlearned model accuracy w.r.t. different model architectures.}
    \resizebox{\linewidth}{!}{%
    \begin{tabular}{lccc}
    \toprule
         & Normal unlearning & Pushing-I & Pushing-II \\
         \midrule
        VGG & 79.3\% & 73.6\% (5.7\%$\downarrow$) & 73.7\% (5.6\%$\downarrow$)\\
        ResNet & 69.4\% & 66.9\% (2.5\%$\downarrow$) & 68.3\% (1.1\%$\downarrow$) \\
        \bottomrule
    \end{tabular}
    }
    
    \label{tab:ab_over_model}
\end{table}

\noindent \textbf{Model Depth.} We study how the depth of the model may affect the effectiveness of Pushing-I and Pushing-II. We set the number of unlearned samples to 400 in the class of ``airplane'' in CIFAR-10. We use the VGG model with the different number of blocks to simulate the models with different depths.  

\autoref{tab:ab_over_depth} shows the accuracy of the unlearned model with different depths under normal unlearning, Pushing-I, and Pushing-II, respectively. As we can see, both Pushing-I and Pushing-II are effective in achieving the goal of over-unlearning on models with different numbers of VGG blocks. When comparing the models with different depths under the same unlearning setting, we find that the utility of the model with more blocks is usually smaller than that of the model with fewer blocks. For example, the unlearned model with 3 blocks has an accuracy of 73.6\% under Pushing-I, while with 5 blocks the model has an accuracy of 65.6\%. Under normal unlearning, the unlearned model with 3 blocks has an accuracy of 79.3\%, while with 5 blocks the unlearned model has an accuracy of 73.3\%. This reminds us that models with deeper depths might be more easily affected by unlearning, either under normal unlearning or malicious unlearning. 

\begin{table}[t!]
    \centering
    \caption{Unlearned model accuracy w.r.t. different model depths.}
    \resizebox{\linewidth}{!}{%
    \begin{tabular}{lccc}
    \toprule
       VGG model  & 3 blocks & 4 blocks & 5 blocks \\
         \midrule
        Normal unlearning &  79.3\%  & 76.5\% & 73.3\%\\ 
        Pushing-I & 73.6\% (5.7\%$\downarrow$) & 72.7\% (3.8\%$\downarrow$) & 65.6\% (7.7\%$\downarrow$)\\
        Pushing-II & 73.7\% (5.6\%$\downarrow$) & 69.5\% (7.1\%$\downarrow$) & 65.7\% (7.6\%$\downarrow$)\\
        \bottomrule
    \end{tabular}
    }
    \label{tab:ab_over_depth}
\end{table}

\noindent \textbf{Class Option.} We study whether Pushing-I and Pushing-II can work when the user has training samples of other classes than the class we have evaluated. We use the VGG model and set the number of unlearned samples to 400. We vary the class of ``bird'', ``horse'', and ``ship'' in CIFAR-10, respectively.

\autoref{tab:ab_over_class} shows the accuracy of the unlearned model under Pushing-I and Pushing-II when the user has the unlearned samples in the class of ``bird'', ``horse'', and ``ship''. As we can see, for all the classes, Pushing-I and Pushing-II can degrade the accuracy of the model by around 2\% to 5\% compared to normal unlearning. The effectiveness of both Pushing-I and Pushing-II in achieving the goal of over-unlearning-I across different classes demonstrates the generalization ability of the advanced pushing methods for over-unlearning.

\begin{table}[t!]
    \centering
    \caption{Unlearned model accuracy w.r.t. different classes.}
    \resizebox{\linewidth}{!}{%
    \begin{tabular}{lccc}
    \toprule
        Base class & Bird & Horse & Ship \\
         \midrule
        Normal unlearning & 76.2\% & 78.4\% & 77.5\%\\
        Pushing-I & 74.1\% (2.1\%$\downarrow$) & 75.6\% (2.8\%$\downarrow$) & 73.5\% (4.0\%$\downarrow$)\\
        Pushing-II & 73.7\% (2.5\%$\downarrow$) & 75.8\% (2.6\%$\downarrow$) & 72.9\% (4.6\%$\downarrow$) \\
        \bottomrule
    \end{tabular}
    }
    
    \label{tab:ab_over_class}
\end{table}

\begin{mdframed}[backgroundcolor=white!10,rightline=true,leftline=true,topline=true,bottomline=true,roundcorner=2mm,everyline=true] 
\textbf{Takeaway 4~} 
\begin{itemize}
\item Pushing-I and Pushing-II are effective across different model architectures and depths, and they are generic to achieve over-unlearning across  different classes.
\item More unlearned samples can result in more effective Pushing-I and Pushing-II for over-unlearning.
\end{itemize}
\end{mdframed}

\section{Discussion}
\noindent \textbf{Naive v.s. Advanced Over-unlearning.}  Even though the blending-based strategy is cost-efficient and
model-irrelevant, we demonstrate that the over-unlearning is conducted in a blind
way. It is only effective in simple classification tasks and hard to generalize to complex tasks with many class
categories or complex patterns. Nevertheless, it is a good motivation for advanced over-unlearning strategies by producing perturbation to push the sample to be close to the decision boundary. 

\noindent \textbf{Difference from Poisoning and Adversarial Attacks.} The blending-based over-unlearning has resemblance to data poisoning, particularly when the subtlety of the label is disregarded. However, a key distinction lies in the fact that our blending scenario does not necessitate the inclusion of a training or re-training procedure, which is typically essential for executing a standard poisoning attack. The pushing pipeline is akin to an adversarial example, but there is a fundamental difference in our objectives. Unlike adversarial attacks which aim to deceive a classifier, our goal is to push the sample closer to the decision boundary. On the other hand, existing advanced techniques for data poisoning attacks and adversarial attacks can be also incorporated into our design. 

\noindent \textbf{Possible Defence.} In our study, we show that over-unlearning can break the trade-off between model performance and unlearning services. To protect against over-unlearning, there are several possible protection strategies the server might leverage. 

\noindent \textit{Hashing as a Possible Defense.} The server can consider hashing techniques \cite{venkatesan2000robust} to verify the authenticity of unlearning requests raised from the users. Specifically, the model owner can hash training samples and send the hash values to the server for storage. The server rejects the malicious unlearning requests if he finds that the hash values of the unlearned sample did not match the stored hash value. However, applying hashing techniques in the MLaaS context is plausible in both theory and practice because of several concerns. The first concern is privacy breach. One of the most significant advantages of isolating the dataset between the model developer and the service provider is to protect the privacy of data contributors. Put differently, the service provider should have no knowledge of the model training sets. Providing hashing results or dataset identification access to service providers may expose the nature of the dataset used locally, leading to potential privacy breaches, such as membership of an individual \cite{shokri2017membership,homer2008resolving} or privacy linkage attack \cite{narayanan2008robust} (\eg re-identify individuals in anonymized datasets). The second concern is false rejections of legitimate unlearning requests. Hashing algorithms designed for ensuring unlearning authenticity may only function effectively when the data samples uploaded by the user and received by the server are identical to the original data samples provided by the user for training the model. If the data sample changes due to compression (eg, PNG to JPEG), network transmission issues, and transcoding, hashing can lead to false rejection of users' legitimate unlearning requests, which may lead to severe consequences for the service provider, including potential legal fines due to GDPR \cite{rosen2011right}. 

\noindent \textit{Membership Inference.} The server can also consider membership inference techniques \cite{shokri2017membership,carlini2022membership,salem2018ml} for authenticity verification because membership inference techniques can identify whether a data sample is a training sample or not. However, membership inference techniques suffer from heavy computational resources for training an inference model and low inference accuracy. Also, they are not very effective on well-generalized models \cite{hu2022membership}. 

\noindent \textit{Other Defences.} The server can use anomaly detection methodologies~\cite{pang2021deep} to scrutinize the submitted unlearning sample further. However, since the proposed over-unlearning is contingent on the user's request, generic anomaly detection is ineffective due to a lack of training data similar to users' submissions. 
One heuristic mitigation is we recommend the service provider to carefully monitor the run-time model performance during deployment. 

As we discussed above, the existing defence techniques have limitations in defending against over-unlearning threats. More advanced adversarial attacks and more robust countermeasures could be incorporated into the over-unlearning design recursively, which provides an avenue for future research. 

\section{Conclusion}
This paper has provided a pioneering exploration into the threats associated with machine unlearning services in the real Machine Learning as a Service (MLaaS) environment. Through a comprehensive investigation, we identified over-unlearning as a significant risk that can compromise the model's utility when malicious unlearning data is submitted by users. We proposed effective strategies for exploiting these risks, requiring only black-box access to the models, thus shedding light on the vulnerabilities of current unlearning methods in MLaaS contexts. Our extensive experiments and comprehensive ablation studies have shown that these strategies can effectively induce over-unlearning across different settings and with various model architectures. These findings underline the critical need to address the highlighted risks, specifically for servers providing MLaaS to maintain their service integrity and ensure compliance with privacy regulations.

As machine unlearning services continue to become increasingly relevant in light of privacy concerns, our research serves as a stepping stone in understanding and mitigating the potential threats posed by these services. Future work should continue this line of inquiry, developing more secure unlearning methods and policies to assure the delicate balance between data privacy, model utility, and service security in MLaaS environments.

\section*{Acknowledgments}
This work is supported in part by Cybersecurity and Quantum Systems group at CSIRO's Data61, Australian Research Council (ARC) DP240103068, as well as CSIRO – National Science Foundation (US) AI Research Collaboration Program. Minhui Xue and Shuo Wang are the corresponding authors of this paper. 
Haojin Zhu is supported by the National Natural Science Foundation of China under Grant No. 62325207.

\bibliographystyle{IEEEtranS}
\bibliography{reference}

\begin{thebibliography}{10}
\providecommand{\url}[1]{#1}
\csname url@samestyle\endcsname
\providecommand{\newblock}{\relax}
\providecommand{\bibinfo}[2]{#2}
\providecommand{\BIBentrySTDinterwordspacing}{\spaceskip=0pt\relax}
\providecommand{\BIBentryALTinterwordstretchfactor}{4}
\providecommand{\BIBentryALTinterwordspacing}{\spaceskip=\fontdimen2\font plus
\BIBentryALTinterwordstretchfactor\fontdimen3\font minus \fontdimen4\font\relax}
\providecommand{\BIBforeignlanguage}[2]{{%
\expandafter\ifx\csname l@#1\endcsname\relax
\typeout{** WARNING: IEEEtranS.bst: No hyphenation pattern has been}%
\typeout{** loaded for the language `#1'. Using the pattern for}%
\typeout{** the default language instead.}%
\else
\language=\csname l@#1\endcsname
\fi
#2}}
\providecommand{\BIBdecl}{\relax}
\BIBdecl

\bibitem{mlaas}
\BIBentryALTinterwordspacing
{Lufthansa Technik}: Keeping airlines flying optimally with {AI-powered} {TechOps} platform {AVIATAR}. [Online]. Available: \url{https://cloud.google.com/customers/lufthansa}
\BIBentrySTDinterwordspacing

\bibitem{meta}
\BIBentryALTinterwordspacing
Meta fined \$1.3 billion for violating {E.U.} data privacy rules. [Online]. Available: \url{https://www.nytimes.com/2023/05/22/business/meta-facebook-eu-privacy-fine.html}
\BIBentrySTDinterwordspacing

\bibitem{googledele}
\BIBentryALTinterwordspacing
User deletion {API} - overview. [Online]. Available: \url{https://developers.google.com/analytics/devguides/config/userdeletion/v3}
\BIBentrySTDinterwordspacing

\bibitem{baumhauer2022machine}
T.~Baumhauer, P.~Sch{\"o}ttle, and M.~Zeppelzauer, ``Machine unlearning: Linear filtration for logit-based classifiers,'' \emph{Machine Learning}, vol. 111, no.~9, pp. 3203--3226, 2022.

\bibitem{bishop2006pattern}
C.~M. Bishop and N.~M. Nasrabadi, \emph{Pattern recognition and machine learning}.\hskip 1em plus 0.5em minus 0.4em\relax Springer, 2006, vol.~4, no.~4.

\bibitem{bourtoule2021machine}
L.~Bourtoule, V.~Chandrasekaran, C.~A. Choquette-Choo, H.~Jia, A.~Travers, B.~Zhang, D.~Lie, and N.~Papernot, ``Machine unlearning,'' in \emph{IEEE Symposium on Security and Privacy (IEEE S\&P)}, 2021, pp. 141--159.

\bibitem{brophy2021machine}
J.~Brophy and D.~Lowd, ``Machine unlearning for random forests,'' in \emph{International Conference on Machine Learning}, 2021, pp. 1092--1104.

\bibitem{brown2020language}
T.~Brown, B.~Mann, N.~Ryder, M.~Subbiah, J.~D. Kaplan, P.~Dhariwal, A.~Neelakantan, P.~Shyam, G.~Sastry, A.~Askell \emph{et~al.}, ``Language models are few-shot learners,'' \emph{Advances in Neural Information Processing Systems}, vol.~33, pp. 1877--1901, 2020.

\bibitem{cao2015towards}
Y.~Cao and J.~Yang, ``Towards making systems forget with machine unlearning,'' in \emph{IEEE Symposium on Security and Privacy (IEEE S\&P)}, 2015, pp. 463--480.

\bibitem{cao2023stylefool}
Y.~Cao, X.~Xiao, R.~Sun, D.~Wang, M.~Xue, and S.~Wen, ``Stylefool: Fooling video classification systems via style transfer,'' in \emph{44th IEEE Symposium on Security and Privacy (IEEE S\&P)}, 2023, pp. 818--835.

\bibitem{carlini2022membership}
N.~Carlini, S.~Chien, M.~Nasr, S.~Song, A.~Terzis, and F.~Tramer, ``Membership inference attacks from first principles,'' in \emph{2022 IEEE Symposium on Security and Privacy (IEEE S\&P)}, 2022, pp. 1897--1914.

\bibitem{carlini2017towards}
N.~Carlini and D.~Wagner, ``Towards evaluating the robustness of neural networks,'' in \emph{2017 IEEE symposium on security and privacy (IEEE S\&P)}, 2017, pp. 39--57.

\bibitem{chen2017zoo}
P.-Y. Chen, H.~Zhang, Y.~Sharma, J.~Yi, and C.-J. Hsieh, ``Zoo: Zeroth order optimization based black-box attacks to deep neural networks without training substitute models,'' in \emph{Proceedings of the 10th ACM Workshop on Artificial Intelligence and Security}, 2017, pp. 15--26.

\bibitem{chen2017targeted}
X.~Chen, C.~Liu, B.~Li, K.~Lu, and D.~Song, ``Targeted backdoor attacks on deep learning systems using data poisoning,'' \emph{arXiv preprint arXiv:1712.05526}, 2017.

\bibitem{coates2011analysis}
A.~Coates, A.~Ng, and H.~Lee, ``An analysis of single-layer networks in unsupervised feature learning,'' in \emph{Proceedings of the 14th International Conference on Artificial Intelligence and Statistics}, 2011, pp. 215--223.

\bibitem{cook1980characterizations}
R.~D. Cook and S.~Weisberg, ``Characterizations of an empirical influence function for detecting influential cases in regression,'' \emph{Technometrics}, vol.~22, no.~4, pp. 495--508, 1980.

\bibitem{de2021critical}
E.~De~Cristofaro, ``A critical overview of privacy in machine learning,'' \emph{IEEE Security \& Privacy}, vol.~19, no.~4, pp. 19--27, 2021.

\bibitem{di2022hidden}
J.~Z. Di, J.~Douglas, J.~Acharya, G.~Kamath, and A.~Sekhari, ``Hidden poison: Machine unlearning enables camouflaged poisoning attacks,'' in \emph{NeurIPS ML Safety Workshop}, 2022.

\bibitem{federal2021ftc}
{Federal Trade Commission}, ``{FTC} report to congress on privacy and security (2021),'' 2021.

\bibitem{golatkar2020eternal}
A.~Golatkar, A.~Achille, and S.~Soatto, ``Eternal sunshine of the spotless net: Selective forgetting in deep networks,'' in \emph{Proceedings of the IEEE/CVF Conference on Computer Vision and Pattern Recognition}, 2020, pp. 9304--9312.

\bibitem{golatkar2020forgetting}
------, ``Forgetting outside the box: Scrubbing deep networks of information accessible from input-output observations,'' in \emph{16th European Conference on Computer Vision}, 2020, pp. 383--398.

\bibitem{goodfellow2014explaining}
I.~J. Goodfellow, J.~Shlens, and C.~Szegedy, ``Explaining and harnessing adversarial examples,'' in \emph{International Conference on Learning Representations}, 2015.

\bibitem{graves2021amnesiac}
L.~Graves, V.~Nagisetty, and V.~Ganesh, ``Amnesiac machine learning,'' in \emph{Proceedings of the AAAI Conference on Artificial Intelligence}, vol.~35, no.~13, 2021, pp. 11\,516--11\,524.

\bibitem{guo2020certified}
C.~Guo, T.~Goldstein, A.~Hannun, and L.~Van Der~Maaten, ``Certified data removal from machine learning models,'' in \emph{Proceedings of the 37th International Conference on Machine Learning}, 2020, pp. 3832--3842.

\bibitem{he2016deep}
K.~He, X.~Zhang, S.~Ren, and J.~Sun, ``Deep residual learning for image recognition,'' in \emph{Proceedings of the IEEE Conference on Computer Vision and Pattern Recognition}, 2016, pp. 770--778.

\bibitem{homer2008resolving}
N.~Homer, S.~Szelinger, M.~Redman, D.~Duggan, W.~Tembe, J.~Muehling, J.~V. Pearson, D.~A. Stephan, S.~F. Nelson, and D.~W. Craig, ``Resolving individuals contributing trace amounts of {DNA} to highly complex mixtures using high-density {SNP} genotyping microarrays,'' \emph{PLOS Genetics}, vol.~4, no.~8, pp. 1--9, 2008.

\bibitem{hu2022membership}
H.~Hu, Z.~Salcic, L.~Sun, G.~Dobbie, P.~S. Yu, and X.~Zhang, ``Membership inference attacks on machine learning: A survey,'' \emph{ACM Computing Surveys (CSUR)}, vol.~54, no. 11s, pp. 1--37, 2022.

\bibitem{imarc2023machine}
{IMARC Group}, ``Machine learning as a service {(MLaaS)} market,'' 2023.

\bibitem{izzo2021approximate}
Z.~Izzo, M.~A. Smart, K.~Chaudhuri, and J.~Zou, ``Approximate data deletion from machine learning models,'' in \emph{International Conference on Artificial Intelligence and Statistics}, 2021, pp. 2008--2016.

\bibitem{kim2022efficient}
J.~Kim and S.~S. Woo, ``Efficient two-stage model retraining for machine unlearning,'' in \emph{Proceedings of the IEEE/CVF Conference on Computer Vision and Pattern Recognition}, 2022, pp. 4361--4369.

\bibitem{kingma2014adam}
D.~P. Kingma and J.~Ba, ``Adam: A method for stochastic optimization,'' in \emph{3rd International Conference for Learning Representations}, 2015.

\bibitem{koh2017understanding}
P.~W. Koh and P.~Liang, ``Understanding black-box predictions via influence functions,'' in \emph{International Conference on Machine Learning}, 2017, pp. 1885--1894.

\bibitem{krizhevsky2009learning}
A.~Krizhevsky and G.~Hinton, ``Learning multiple layers of features from tiny images,'' 2009.

\bibitem{uk}
\BIBentryALTinterwordspacing
B.~Levin and L.~Downes. Who is going to regulate {AI}? [Online]. Available: \url{https://hbr.org/2023/05/who-is-going-to-regulate-ai.}
\BIBentrySTDinterwordspacing

\bibitem{liu2017delving}
Y.~Liu, X.~Chen, C.~Liu, and D.~Song, ``Delving into transferable adversarial examples and black-box attacks,'' in \emph{International Conference on Learning Representations}, 2017.

\bibitem{marchant2022hard}
N.~G. Marchant, B.~I. Rubinstein, and S.~Alfeld, ``Hard to forget: Poisoning attacks on certified machine unlearning,'' in \emph{Proceedings of the AAAI Conference on Artificial Intelligence}, vol.~36, no.~7, 2022, pp. 7691--7700.

\bibitem{narayanan2008robust}
A.~Narayanan and V.~Shmatikov, ``Robust de-anonymization of large sparse datasets,'' in \emph{IEEE Symposium on Security and Privacy (IEEE S\&P)}, 2008, pp. 111--125.

\bibitem{narayanan2021efficient}
D.~Narayanan, M.~Shoeybi, J.~Casper, P.~LeGresley, M.~Patwary, V.~Korthikanti, D.~Vainbrand, P.~Kashinkunti, J.~Bernauer, B.~Catanzaro \emph{et~al.}, ``Efficient large-scale language model training on gpu clusters using megatron-lm,'' in \emph{Proceedings of the International Conference for High Performance Computing, Networking, Storage and Analysis}, 2021, pp. 1--15.

\bibitem{neel2021descent}
S.~Neel, A.~Roth, and S.~Sharifi-Malvajerdi, ``Descent-to-delete: Gradient-based methods for machine unlearning,'' in \emph{Algorithmic Learning Theory}, 2021, pp. 931--962.

\bibitem{pang2021deep}
G.~Pang, C.~Shen, L.~Cao, and A.~V.~D. Hengel, ``Deep learning for anomaly detection: A review,'' \emph{ACM Computing Surveys (CSUR)}, vol.~54, no.~2, pp. 1--38, 2021.

\bibitem{papernot2016limitations}
N.~Papernot, P.~McDaniel, S.~Jha, M.~Fredrikson, Z.~B. Celik, and A.~Swami, ``The limitations of deep learning in adversarial settings,'' in \emph{IEEE European Symposium on Security and Privacy (EuroS\&P)}, 2016, pp. 372--387.

\bibitem{pardau2018california}
S.~L. Pardau, ``The {California} consumer privacy act: Towards a {European-style} privacy regime in the {United States},'' \emph{Journal of Technology Law \& Policy}, vol.~23, p.~68, 2018.

\bibitem{rosen2011right}
J.~Rosen, ``The right to be forgotten,'' \emph{Stanford Law Review}, vol.~64, p.~88, 2011.

\bibitem{salem2018ml}
A.~Salem, Y.~Zhang, M.~Humbert, P.~Berrang, M.~Fritz, and M.~Backes, ``{ML-leaks}: Model and data independent membership inference attacks and defenses on machine learning models,'' in \emph{Network and Distributed Systems Security (NDSS) Symposium}, 2019.

\bibitem{samek2017explainable}
W.~Samek, T.~Wiegand, and K.-R. M{\"u}ller, ``Explainable artificial intelligence: Understanding, visualizing and interpreting deep learning models,'' \emph{arXiv preprint arXiv:1708.08296}, 2017.

\bibitem{shannon2001mathematical}
C.~E. Shannon, ``A mathematical theory of communication,'' \emph{ACM SIGMOBILE Mobile Computing and Communications Review}, vol.~5, no.~1, pp. 3--55, 2001.

\bibitem{shmueli2023machine}
G.~Shmueli, P.~C. Bruce, K.~R. Deokar, and N.~R. Patel, \emph{Machine Learning for Business Analytics: Concepts, Techniques, and Applications with Analytic Solver Data Mining}.\hskip 1em plus 0.5em minus 0.4em\relax John Wiley \& Sons, 2023.

\bibitem{shokri2017membership}
R.~Shokri, M.~Stronati, C.~Song, and V.~Shmatikov, ``Membership inference attacks against machine learning models,'' in \emph{IEEE Symposium on Security and Privacy (IEEE S\&P)}, 2017, pp. 3--18.

\bibitem{simonyan2014very}
K.~Simonyan and A.~Zisserman, ``Very deep convolutional networks for large-scale image recognition,'' in \emph{3rd International Conference for Learning Representations}, 2015.

\bibitem{tarun2023fast}
A.~K. Tarun, V.~S. Chundawat, M.~Mandal, and M.~Kankanhalli, ``Fast yet effective machine unlearning,'' \emph{IEEE Transactions on Neural Networks and Learning Systems}, 2023.

\bibitem{thudi2022unrolling}
A.~Thudi, G.~Deza, V.~Chandrasekaran, and N.~Papernot, ``Unrolling sgd: Understanding factors influencing machine unlearning,'' in \emph{IEEE 7th European Symposium on Security and Privacy (EuroS\&P)}, 2022, pp. 303--319.

\bibitem{venkatesan2000robust}
R.~Venkatesan, S.-M. Koon, M.~H. Jakubowski, and P.~Moulin, ``Robust image hashing,'' in \emph{Proceedings 2000 International Conference on Image Processing (Cat. No. 00CH37101)}, vol.~3, 2000, pp. 664--666.

\bibitem{wang2023publicheck}
S.~Wang, S.~Abuadbba, S.~Agarwal, K.~Moore, R.~Sun, M.~Xue, S.~Nepal, S.~Camtepe, and S.~Kanhere, ``Publiccheck: Public watermarking verification for deep neural networks,'' in \emph{44th IEEE Symposium on Security and Privacy (IEEE S\&P)}, 2023.

\bibitem{wang2004image}
Z.~Wang, A.~C. Bovik, H.~R. Sheikh, and E.~P. Simoncelli, ``Image quality assessment: from error visibility to structural similarity,'' \emph{IEEE Transactions on Image Processing}, vol.~13, no.~4, pp. 600--612, 2004.

\bibitem{warnecke2023machine}
A.~Warnecke, L.~Pirch, C.~Wressnegger, and K.~Rieck, ``Machine unlearning of features and labels,'' in \emph{Network and Distributed System Security Symposium (NDSS)}, 2023.

\bibitem{wu2020deltagrad}
Y.~Wu, E.~Dobriban, and S.~Davidson, ``Deltagrad: Rapid retraining of machine learning models,'' in \emph{International Conference on Machine Learning}, 2020, pp. 10\,355--10\,366.

\bibitem{xu2023machine}
H.~Xu, T.~Zhu, L.~Zhang, W.~Zhou, and P.~S. Yu, ``Machine unlearning: A survey,'' \emph{ACM Computing Surveys}, 2023.

\bibitem{yue2022gradient}
K.~Yue, R.~Jin, C.-W. Wong, D.~Baron, and H.~Dai, ``Gradient obfuscation gives a false sense of security in federated learning,'' in \emph{{USENIX} Security Symposium}, 2023.

\bibitem{zhang2018unreasonable}
R.~Zhang, P.~Isola, A.~A. Efros, E.~Shechtman, and O.~Wang, ``The unreasonable effectiveness of deep features as a perceptual metric,'' in \emph{Proceedings of the IEEE Conference on Computer Vision and Pattern Recognition}, 2018, pp. 586--595.

\end{thebibliography}

\balance
\section*{Appendix}
\setcounter{section}{0}
\renewcommand{\thesection}{\Alph{section}}
\section{Approximate Unlearning}
\label{appendix:unlearningeq}

There are mainly two kinds of approximate unlearning methods, which are detailed introduced as follows. Let $\bm{\theta}^*$ be the trained model. Let $\mathcal{D}_{\textrm{train}}=\mathcal{D}_{u} \cup \mathcal{D}_{r}$, where $\mathcal{D}_{u}$ is the dataset that contains the unlearned data and $\mathcal{D}_{r}$ is the dataset that contains the remaining data. Specifically, the first kind of approximate unlearning method~\cite{guo2020certified,izzo2021approximate} leverages the influence function to calculate the influence of the unlearned data on the model parameters so that the trained model can apply a Newton step to remove the influence for obtaining the unlearned model:
\begin{equation}
    \bm{\theta}_{u}^*=\bm{\theta}^*+\mathcal{H}_{\bm{\theta}}^{-1}\Delta_{\bm{\theta}}(\bm{Z}), 
\end{equation}
where $\bm{Z}$ is the unlearned data. $\mathcal{H}_{\bm{\theta}}$ and $\Delta_{\bm{\theta}}(\bm{Z})$ is defined as:
\begin{equation}
    \mathcal{H}_{\bm{\theta}}=\nabla^2\mathcal{L}(\bm{\theta}^*,\mathcal{D}_{r}),
\end{equation}
\begin{equation}
    \Delta_{\bm{\theta}}(\bm{Z})= \displaystyle\sum_{{\bm{z}} \in {\bm{Z}}} \nabla \mathcal{L}({\bm{z}},\bm{\theta}^{*}).
\end{equation}
Here, $\mathcal{H}_{\bm{\theta}}$ is the Hessian matrix of the loss function $\mathcal{L}(.,\mathcal{D}_{r})$ at $\bm{\theta}^*$. $\mathcal{H}_{\bm{\theta}}^{-1}$ is the inverse of $\mathcal{H}_{\bm{\theta}}$. The term $\mathcal{H}_{\bm{\theta}}^{-1}\Delta_{\bm{\theta}}(\bm{Z})$ can be interpreted as the influence of $\mathcal{D}_{u}$ on the model parameter  $\bm{\theta}^*$~\cite{koh2017understanding}. The advantage of the first kind of approximate unlearning method is that one-step Newton update is enough for removing the contribution of $\mathcal{D}_{u}$. However, the drawbacks are also obvious: \textit{i)} the construction of the inverse Hessian matrix can be difficult for large-scale deep models; \textit{ii)} the $\mathcal{D}_{r}$ might not be applicable in cases where $\mathcal{D}_{\textrm{train}}$ is not stored or deleted.

The second kind of unlearning method~\cite{thudi2022unrolling,wu2020deltagrad,neel2021descent} calculates the gradients of the unlearned data contributed to the trained model during the training process. Then, to unlearn the data $\mathcal{D}_{u}$, the trained model $\bm{\theta}^*$ is updated by adding back these gradients to approximate the model $\bm{\theta}$ that is retrained on $\mathcal{D}_{r}$. A state-of-the-art gradient-based unlearning method~\cite{warnecke2023machine} is as follows:
\begin{equation}\label{equ:un1}
    \bm{\theta}^{*}_{u} \approx \bm{\theta}^{*} + \Delta(\bm{Z},\hat{\bm{Z}}),
\end{equation}
where
\begin{equation}\label{equ:un2}
    \Delta(\bm{Z},\hat{\bm{Z}})=-\tau \Bigl( \displaystyle\sum_{\hat{\bm{z}} \in \hat{\bm{Z}}} \nabla_{\bm{\theta}} \mathcal{L}(\hat{\bm{z}},\bm{\theta}^{*}) - \displaystyle\sum_{{\bm{z}} \in {\bm{Z}}} \nabla_{\bm{\theta}} \mathcal{L}({\bm{z}},\bm{\theta}^{*}) \Bigl).
\end{equation}
Here, $\hat{\bm{Z}}$ is the perturbed version of the unlearned data $\bm{Z}$, and $\tau$ is a small constant of the unlearning rate. The high-level intuition of this unlearning method~\cite{warnecke2023machine} is to overwrite $\bm{Z} \in \mathcal{D}_u$ from $\bm{\theta}^*$ by using the perturbed data $\hat{\bm{Z}}$. Thus, $\Delta(\bm{Z},\hat{\bm{Z}})$ can be interpreted as the direction that moves the model $\bm{\theta}$ to remove the information contained in $\mathcal{D}_u$. An advantage of this method~\cite{warnecke2023machine} is that it only requires to access $\mathcal{D}_{u}$, which makes it practical in many cases where only $\mathcal{D}_{u}$ is applicable.

\begin{table*}[t]
    \centering
    \caption{The overall accuracy of the model before unlearning, with normal unlearning, with the blending method on CIFAR-10, CIFAR-100, and STL-10.}
    \resizebox{1\textwidth}{!}{%
    \begin{tabular}{lcccccc}
    \toprule
        &  & & \multicolumn{2}{c}{Test accuracy when unlearning 10\% data of a class} & \multicolumn{2}{c}{Test accuracy when unlearning 50\% data of a class} \\
       \cmidrule(lr){4-5}  \cmidrule(l){6-7}
        Dataset & Training accuracy & Test accuracy & Normal unlearn & Blending & Normal unlearn & Blending  \\
        \midrule
       CIFAR-10 & 81.5\% & 79.8\% & 79.3\% & 73.2\% & 78.7\% & 65.0\%  \\
       CIFAR-100 & 76.3\%  & 51.1\% & 51.1\% & 50.9\%  & 49.7\% & 49.2\%	 \\
       STL-10 & 96.3\%& 56.6\% & 56.6\% & 56.5\% & 50.2\% &  49.0\%  \\
       \bottomrule
    \end{tabular}
}
    
    \label{tab:acc_overall_blend}
\end{table*}

\begin{table*}[t]
    \centering
    \caption{The overall accuracy of the model before unlearning, with normal unlearning, with Pushing-I, and with Pushing-II on CIFAR-10, CIFAR-100, and STL-10 when the server uses fine-tuning as the unlearning method.}
    \resizebox{1\textwidth}{!}{%
    \begin{tabular}{lcccccccc}
    \toprule
         &  & & \multicolumn{3}{c}{Test accuracy when unlearning 10\% data of a class} & \multicolumn{3}{c}{Test accuracy when unlearning 50\% data of a class} \\
       \cmidrule(lr){4-6}  \cmidrule(l){7-9}
        Dataset & Training accuracy & Test accuracy & Normal unlearning & Pushing-I & Pushing-II & Normal unlearning & Pushing-I & Pushing-II  \\
       \midrule
       CIFAR-10 &  81.5\% & 79.8\% & 80.6\% & 79.8\% (0.8\%$\downarrow$) & 79.6\%(1.0\%$\downarrow$) &  78.9\% & 75.9\% (3.0\%$\downarrow$) & 75.8\% (0.1\%$\downarrow$) \\
       CIFAR-100 & 76.3\% & 51.1\% & 51.3\% & 50.8\% (0.5\%$\downarrow$) & 51.0\% (0.3\%$\downarrow$) & 51.3\% & 50.8\% (0.5\%$\downarrow$) & 50.7\% (0.6\%$\downarrow$) \\
       STL-10 & 96.3\%& 56.6\% & 58.3\% & 56.6\% (1.7\%$\downarrow$) & 56.9\% (1.4\%$\downarrow$) & 56.8\% &  56.1\% (0.7\%$\downarrow$) & 56.0\% (0.8\%$\downarrow$) \\
       \bottomrule
    \end{tabular}
}
    
    \label{tab:acc_pushing_overall_finetune}
\end{table*}

\begin{table}[t]
\centering
\caption{Quantitative stealthiness analysis on CIFAR-10}
\resizebox{1\linewidth}{!}{%
\begin{tabular}{@{}lcc@{}}
\toprule
& SSIM (mean $\pm$ std) & LPIPS (mean $\pm$ std)\\
\midrule
Pushing-I & 0.9748 $\pm$ 0.0013 & 0.0339 $\pm$ 0.0015\\

Pushing-II & 0.9691 $\pm$ 0.0026 & 0.0374 $\pm$ 0.0025\\

\bottomrule
\end{tabular}
}

\label{tab:ab_decp_blend}
\end{table}

\section{Experimental Settings}
\label{appendix:settigns}
\noindent \textbf{Dataset Division.} For all the datasets, we use 80\% of the training data to train the model and use 20\% of the training data as the validate data to prevent the model from overfitting. We aim to demonstrate the effectiveness of our methods on well-generalized models. For STL-10, to align the percentage of training data, validation data, and test data with CIFAR-10 and CIFAR-100, we select 1,000 test images from the 8,000 test samples and report the test accuracy of the model on these 1,000 test samples. 

\noindent \textbf{Models.} We use two VGG models and one ResNet model in our experiments. The three deep models are with different depths and architectures to mimic the server's different normal business tasks. The first deep learning model is composed of three VGG blocks~\cite{simonyan2014very} (consisting of six convolutional layers) and two dense layers. This model utilizes 128 convolutional filters, a kernel size of 3 × 3, a pooling size of 2 × 2, and the ReLU activation function. Our experiments are mainly conducted on the first model. The second model follows the same configuration but is deeper, featuring five VGG blocks encompassing ten convolutional layers. In contrast, the final model replaces the VGG blocks from the first model with Resnet blocks~\cite{he2016deep}. It incorporates one convolutional layer, three Resnet blocks, and two dense layers. The first convolutional layer has a filter size of 64 and a kernel size of 3 × 3. The first two Resnet blocks comprise two convolutional layers with 64 filters and a 3 × 3 kernel size, utilizing the ReLU activation function. The last Resnet block has a similar structure to the preceding blocks, except it employs 128 filters and includes an additional down sampling block. The down sampling block consists of one convolutional layer with a 1 × 1 kernel size and 2 strides size to concatenate the output results of the second and last Resnet blocks. 

All three networks are trained using the Adam optimizer~\cite{kingma2014adam} with a learning rate at 0.001. To mitigate the issue of model overfitting, the validation set is used for terminating the training process at the lowest validation loss value.

\section{Effectiveness of Over-unlearning on the Fine-tuning based Unlearning Method}
\label{appendix:fine}
Using fine-tuning as an unlearning method is to confuse the model’s understanding of the unlearned samples by replacing their original labels with random labels. Thus, by fine-tuning the model on such relabeled samples, the model cannot output the correct prediction of the unlearned data. Because the blending method is not reliable for over-unlearning, we mainly conduct experiments using the advanced over-unlearning methods of Pushing-I and Pushing-II.

\autoref{tab:acc_pushing_overall_finetune} shows the overall accuracy of the model on all classes. As we can see, Pushing-I and Pushing-II can degrade the unlearned model's accuracy compared to normal unlearning, although the degradation is small. This indicates the effectiveness of Pushing-I and Pushing-II for fine-tuning based unlearning method. The degradation is not obvious as the case on the gradient-based unlearning method is because the well-designed additional information embedded into the unlearned samples is harmed by the randomly reassigned labels. It is essentially very difficult to achieve over-unlearning on fine-tuning. However, fine-tuning cannot guarantee the information of the unlearned data is removed from the parameters of the model, which can make it less attractive for the server.

\section{Stealthiness of Modified Unlearned Samples} 
\label{appendix:similarity}
Following the methodology of recent works to perceive stealthiness~\cite{cao2023stylefool,wang2023publicheck,yue2022gradient}, we quantitatively evaluate the stealthiness of modified unlearned samples with widely used perceptual metrics, such as Structural Similarity Index Measure (SSIM~\cite{wang2004image}), 
and Learned Perceptual Image Patch Similarity  (LPIPS~\cite{zhang2018unreasonable}).
The SSIM index is a decimal value between $-1$ and $1$, where $1$ indicates perfect similarity, $0$ indicates no similarity, and $-1$ indicates perfect anti-correlation. The closer the SSIM value to 1, the more similar the image pair is.  
LPIPS judges the perceptual similarity between two images through computing the similarity between the activations of two images for some pre-defined network. Lower LPIPS values indicate higher similarity. The smaller, the more similar.

Particularly, we randomly selected 50 pairs of original samples and corresponding modified unlearned samples from the dataset CIFAR-10. The modified unlearned samples are generated by strategies Pushing-I and Pushing-II, respectively. We calculate the mean and standard deviation of SSIM and LPIPS.
According to \autoref{tab:ab_decp_blend}, the experimental results show high similarities between modified unlearning samples and original samples, indicating commendable stealthiness performance.

\end{document}